\UseRawInputEncoding
    \documentclass[prresearch,twocolumn,superscriptaddress,floatfix,nopacs,nofootinbib,longbibliography]{revtex4-2}

    \usepackage{graphicx,amsfonts,amssymb,amsmath,hyperref}
    \usepackage{lmodern}
    \usepackage{xcolor}
    \usepackage{pgfplots}
    \usepackage{pgfplotstable}
    \pgfplotsset{compat=1.18}
    \usepgfplotslibrary{groupplots}

    \definecolor{mvcrose}{HTML}{F01438}
    \definecolor{mvcblue}{HTML}{33BBFF}
    \definecolor{mvcpurple}{HTML}{A455FF}
    \definecolor{mvcgray900}{HTML}{21262B}
    \definecolor{mvcgray500}{HTML}{D1D6DB}
    \definecolor{mvcgray700}{HTML}{5E666E}

    \newif\ifhyper
    \hypertrue
    \ifhyper
    \hypersetup{
    citecolor = {red},
    colorlinks = {true},
    urlcolor = {blue}
    }
    \fi

    \def\bra#1{\langle#1\vert}
    \def\ket#1{\vert#1\rangle}

    \let\oldpm\pm
    \renewcommand{\pm}{\,\mathord{\oldpm}\,}
    \let\oldtimes\times
    \renewcommand{\times}{\,\mathord{\oldtimes}\,}
    
\begin{document}

    \title{Multi-head Quantum Neural Networks:\texorpdfstring{ \\}{ }Replacing Transformer Blocks with NISQ Modules inside a 30-Billion-Parameter LLM}

    \author{Maforikan Amoussou}
    \affiliation{Multiverse Computing, Toronto, Ontario, Canada}
    \affiliation{The Edward S. Rogers Sr. Department of Electrical and Computer Engineering - University of Toronto, Toronto, Ontario, Canada}

    \author{Borja Aizpurua}
    \affiliation{Multiverse Computing, San Sebasti\'{a}n, Spain}
    \affiliation{Department of Basic Sciences, Tecnun - University of Navarra, San Sebasti\'an, Spain}

    \author{Sukhbinder Singh}
    \affiliation{Multiverse Computing, Toronto, Ontario, Canada}

    \author{Rom\'{a}n Or\'{u}s}
    \affiliation{Multiverse Computing, San Sebasti\'{a}n, Spain}
    \affiliation{Donostia International Physics Center, San Sebasti\'an, Spain}
    \affiliation{Ikerbasque Foundation for Science, Bilbao, Spain}

    \author{Hans-Arno Jacobsen}
    \affiliation{The Edward S. Rogers Sr. Department of Electrical and Computer Engineering - University of Toronto, Toronto, Ontario, Canada}

    \begin{abstract}

    We introduce the multi-head quantum neural network (QNN), a quantum module compatible with noisy intermediate-scale quantum (NISQ) devices
    that can replace transformer blocks inside pre-trained large language
    models (LLMs) with little loss in accuracy and with far fewer parameters than the blocks it replaces. Encoding a model's $D$ hidden neurons into a qubit register is the obstacle: amplitude encoding scales poorly in circuit depth and angle encoding in qubit count, so neither is practical at production-scale width. We instead partition the hidden space into $k$ disjoint heads,
    angle-encode each into its own $n_q$-qubit register, process the heads in parallel,
    and couple them through constrained classical linear layers, the \emph{mixers}. Performing
    surgery on Qwen3-30B-A3B-Instruct-2507, we show that a single multi-head QNN with $145\,281$
    trainable parameters ($D=2048$, $k=32$, $n_q=8$, $L=4$) can replace a contiguous span of
    eight transformer blocks, displacing $16.33\%$ of the model's parameters. Ablations
    show that, for a single replaced block, the quantum heads (circuits followed by a learned readout), and not the classical mixers alone, recover the displaced function, whereas for fused spans the mixers alone come within $1$--$3\%$ of the full module in perplexity. Whether this benefit is specific to trained, entangling quantum circuits is not established: controls that switch the entangler off (two seeds), or freeze the circuit at its initial or at uniformly random angles (one seed), beat the full module on held-out perplexity, and a parameter-matched classical multilayer perceptron remains a few percent ahead of the full module. We
    train classically with exact statevector
    simulation, then execute the quantum heads of our fusion checkpoints on a $156$-qubit
    IBM Heron processor, showing that NISQ-scale quantum modules can be installed in a production-scale LLM and run on existing hardware.

    \end{abstract}

    \maketitle

    \section{Introduction}
    \label{sec:intro}

    Quantum machine learning (QML) has been developed almost entirely at small scale. QML models
    are trained on low-dimensional inputs and benchmark tasks of a few dozen
    features~\cite{huang2021power,caro2022generalization}, while modern classical machine learning models 
    carry hidden states of thousands of dimensions inside networks
    of hundreds of billions of weights. QML work that has reached large language models (LLMs) at all
    has either adapted classical layers (converting trained
    weights into quantum-executable form through tensor-network
    disentanglers~\cite{aizpurua2024qllm,aizpurua2025classical}, or training unitary
    adapters on a converted model, which changes its function, and executing them on quantum
    hardware~\cite{aizpurua2024qllm,aizpurua2026quantumenhancedlargelanguagemodels}),
    or added quantum components to small language models, as in the hybrid quantum transformer HyQuT~\cite{kong2025hyqut}, which replaces about $10\%$ of the parameters of a $150$-million-parameter model. Other groups have pursued quantum-assisted LLMs along related lines, including fine-tuning~\cite{kong2025quantumenhanced}, distillation~\cite{li2025quantum,liu2025quantum} and natural-language generation~\cite{chen2025quantumenhanced}. Quantum self-attention layers have been proposed for small text and image classifiers~\cite{li2022qsann,zhao2023qksan} and inside hybrid transformers~\cite{chen2025qasa}.

    In this paper, we introduce a quantum module, the \emph{multi-head quantum neural network} (QNN), a straightforward way of encoding an activation vector of the $D$ input neurons into a quantum state and then
    processing that state with a QNN whose outputs, with one upload per feature, are multilinear trigonometric polynomials of the encoded features with one learned frequency per feature. We partition $D$ neurons into $k$ heads of size $d_h \ll D$, dense angle-encode~\cite{larose2020robust}
    each segment into its own small register, and process the heads in parallel with
    independent QNNs.

    We then demonstrate one application of multi-head QNNs: they can replace
    transformer blocks inside Qwen3-30B-A3B-Instruct-2507~\cite{qwen3,qwen3_2507_card} (hereafter Qwen3-30B-A3B), a production-scale LLM. The module is not
    specific to this application, and QNNs of this kind are widely studied as machine-learning
    models~\cite{biamonte2017qml,mitarai2018qcl,benedetti2019pqc,cerezo2021vqa,schuldkilloran2019feature,havlicek2019supervised}; here it approximates the function of a displaced block. It embeds narrow, shallow quantum circuits inside a wide classical network, and the growing width of the host model is absorbed by the number of heads
    rather than by the size of any one register.

    It is not obvious that a transformer block should be replaceable by a quantum
    component at all. The reason to think it might be is redundancy. A trained transformer
    tolerates the outright removal of entire blocks with surprisingly little
    loss~\cite{men2024shortgpt,gromov2024unreasonable}, trained lightweight modules can stand in for the removed layers~\cite{chen2025streamline}, and the same premise underwrites
    pruning~\cite{frantar2023sparsegpt}, quantization~\cite{frantar2023gptq}, and low-rank or tensor-network
    factorization~\cite{tomut2024compactifai}: the function of a block can often
    be supplied by something far smaller than the block itself. Nothing then requires that
    what supplies it be a classical neural network.

    We demonstrate that a multi-head QNN can indeed effectively replace a span of eight blocks (the widest we tested that stays within about $1\%$ of the unmodified model in perplexity on both held-out sets), displacing 16.33\% of the model's parameters. We execute the quantum heads of the five fusion checkpoints on a 156-qubit IBM Heron quantum processor for the first prompt token, and for $N=8$ at every decoding step of two $15$-token generations, with the rest of the model on GPUs (Sec.~\ref{sec:hardware}). These runs show that the circuits
    execute as compiled on hardware that exists today. To our knowledge, this is the first replacement of transformer-block function by a quantum module in an LLM of this size. We claim no quantum advantage and nothing here is a new capability
    for the LLM itself; a parameter-matched classical multilayer perceptron (MLP) is ahead of it in perplexity, by a few percent, in all but one of our comparisons. What is new is the application: a demonstration that NISQ-scale circuits
    can take on part of a production model's function at all, where the classical analogue, replacing pruned layers with a trained lightweight module~\cite{chen2025streamline}, is established.

    Our principal motivation is the engineering question of whether a large-scale hybrid
    quantum--classical LLM can be built and run, and this is the only question the paper addresses. A further, more
    forward-looking motivation is \emph{parameter efficiency}: a circuit layer on $n_q$
    qubits stores $\mathcal{O}(n_q)$ parameters while acting on a $2^{n_q}$-dimensional space,
    so a circuit that supplies the function of a transformer block may do so with far
    fewer stored weights. At NISQ scale this concerns parameter counts alone, since
    circuits of the size we use can be simulated exactly on classical hardware. Larger
    circuits would leave that regime, though whether a useful window exists there remains
    unclear: trainability degrades as circuits
    grow~\cite{mcclean2018barren,cerezo2021cost,ragone2024lie,larocca2025barren}, and the
    structural restrictions that mitigate this may also preserve classical
    simulability~\cite{cerezo2025simulability}.

    Our contributions are:
    \begin{itemize}
    \setlength{\itemsep}{0pt}\setlength{\parskip}{0pt}\setlength{\topsep}{2pt}
    \item the multi-head QNN, whose width scales with the number of heads rather than with any one register (Sec.~\ref{sec:method});
    \item replacement of one block, or of spans of up to sixteen blocks, of Qwen3-30B-A3B by one such module, against identity-skip, ablations and a parameter-matched MLP, with near-parity on held-out perplexity up to eight fused blocks (Sec.~\ref{sec:results});
    \item execution of the trained heads of the five fusion checkpoints on a $156$-qubit IBM Heron processor, with per-observable deviations from statevector simulation measured for $N=8$ over full generations (Sec.~\ref{sec:hardware}); and
    \item an analysis of the learned readout and a statement of what the evidence does not show, including circuit controls showing that the benefit does not require trained or entangled circuits, and the controls it still lacks (Sec.~\ref{sec:discussion}).
    \end{itemize}

    Sec.~\ref{sec:method} describes the architecture, Sec.~\ref{sec:results} the replacement experiments and the hardware execution (Sec.~\ref{sec:hardware}), Sec.~\ref{sec:discussion} the comparison with a parameter-matched (classical) MLP and the limits of the replacement, and Sec.~\ref{sec:conclusions} concludes. Complete data and protocols are given in the appendices.

    \section{Multi-head Quantum Neural Networks}
    \label{sec:method}

    Amplitude encoding is the most qubit-efficient way of encoding data into a quantum circuit, packing $D$ amplitudes into
    $\log_2 D$ qubits.
    The cost, however, is hidden in state preparation: loading a generic $D$-dimensional
    vector into $\log_2 D$ qubits requires circuits of depth $\mathcal{O}(D)$~\cite{shende2006synthesis,plesch2011state}, far beyond what NISQ hardware can execute coherently.

    Angle encoding inverts the trade-off.
    Each activation is written into the rotation angle of a single-qubit gate, so state
    preparation costs a single shallow layer of gates and remains well within the
    coherence budget of present-day devices.
    The width, however, now grows linearly rather than logarithmically with the number of
    encoded features, and, given current frontier models' hidden dimensions, is far
    outside the NISQ regime.

    Neither is practical at the scale of a contemporary LLM. A multi-head QNN instead distributes the hidden state across many small QNNs evaluated in
    parallel, each using dense angle encoding (DAE) together with layerwise
    uploads~\cite{barrue2024layered} across $L$ circuit layers, so that a single circuit
    ingests $2n_qL$ features.
    While the number of qubits available to one QNN is constrained by hardware and by the
    onset of barren plateaus~\cite{mcclean2018barren,larocca2025barren}, no such
    constraint applies to the number of independent QNNs evaluated in parallel.
    We accordingly instantiate $k$ heads such that $D = 2n_qLk$, partitioning the
    $D$-dimensional feature vector $\mathbf{x}$ into $k$ disjoint subspaces and assigning
    one QNN to apply a nonlinear transformation to each.
    Here $n_q$ and $L$ are chosen to keep the resulting circuits executable on current
    hardware at tolerable error rates, and $k$ absorbs the remaining width of the model.
    We write $d_h = 2n_qL$ for the width of a single head's subspace (for a single upload pass; variants with several passes re-upload features, see Appendix~\ref{annex:model}).

    Both register width and circuit depth constrain this choice. On the hardware side, error accumulates with both: more qubits means more physical
    error channels held coherent at once, and more layers means a longer two-qubit chain in
    which those errors compound.
    On the optimization side, both enter the barren-plateau
    literature~\cite{mcclean2018barren,cerezo2021cost,ortizmarrero2021entanglement,ragone2024lie,larocca2025barren},
    where the variance of the loss gradient decays exponentially in the system size and is
    further suppressed as the circuit approaches the expressivity of a random unitary at
    increasing depth.
    Depth, set by the $(n_q-1)L$ two-qubit rotations of a head, dominates the
    compiled cost we report in Sec.~\ref{sec:hardware} and is
    the more limiting of the two in getting a circuit to execute at
    useful fidelity.

    The construction partitions along width, holding $n_q$ small and
    absorbing the model's growth into $k$, rather than deepening or widening a single
    register; the open $R_{zz}$ chain and fractional-gate compilation (Sec.~\ref{sec:hardware}) keep each register cheap to execute.

    \subsection{Mixers}

    The construction above has one immediate limitation.
    Because the heads are mutually independent, they cannot exchange information, and the
    map they collectively realize is block-separable in the
    $D$-dimensional feature vector $\mathbf{x}$: each of the $k$ blocks is transformed independently.
    Attention heads inside a multi-head attention block likewise act on separate subspaces
    of the activation space, but they are coupled by dense projections into and out of
    those subspaces~\cite{vaswani2017attention}; a purely block-separable quantum module,
    with no communication between heads, would therefore be poorly matched to the
    computation it replaces.

\begin{figure*}[t]
        \centering
        \includegraphics[width=15cm]{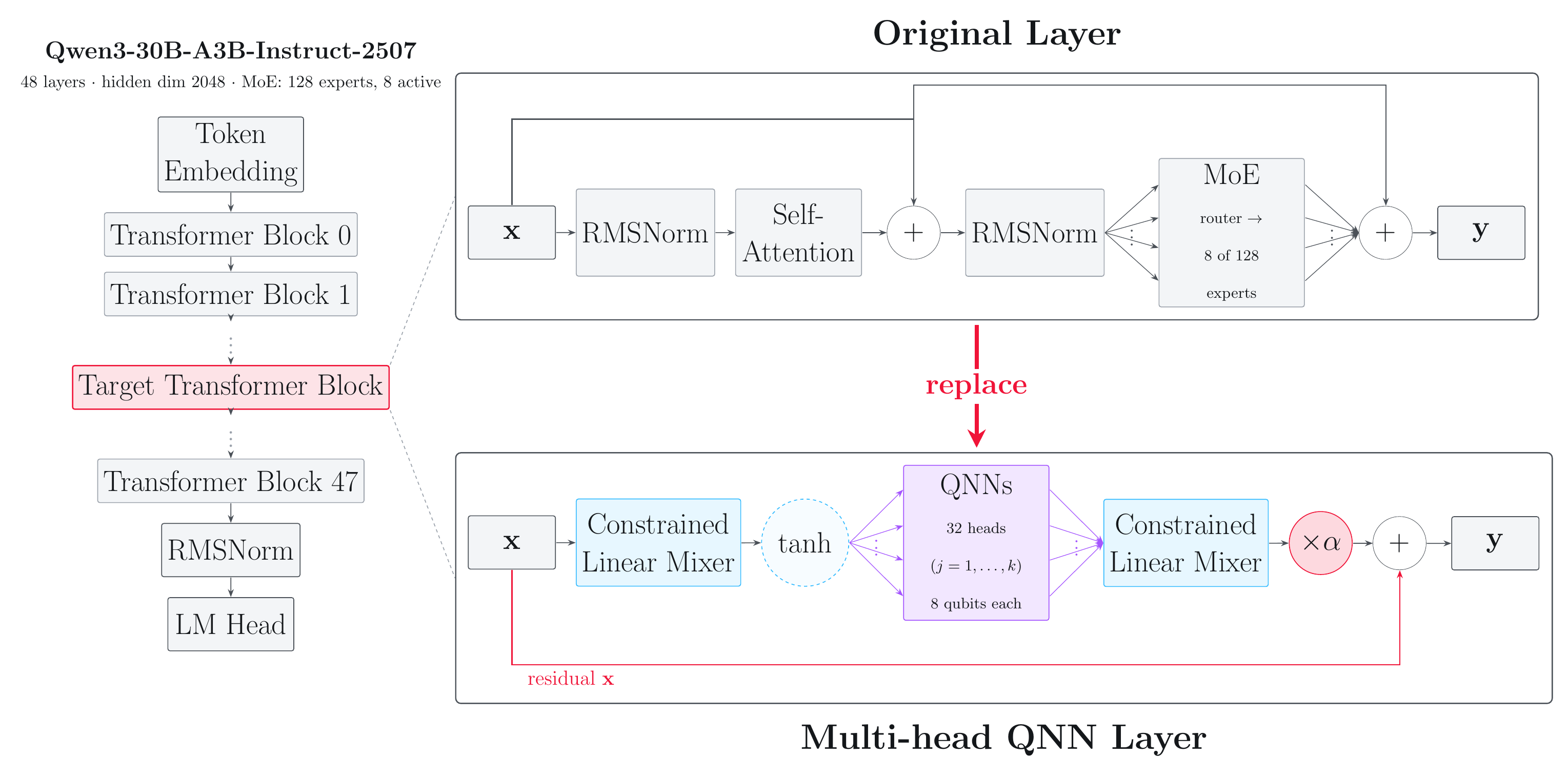}
        \caption{%
        Integration of the multi-head QNN block into Qwen3-30B-A3B, a 48-layer mixture-of-experts (MoE) model with hidden dimension $D=2048$.
        \textbf{Left:} a standard transformer block, consisting of self-attention preceded by RMSNorm~\cite{zhang2019rmsnorm} and a sparse MoE feed-forward network that routes each token to $8$ of $128$ experts, each wrapped in a residual connection.
        \textbf{Right:} the multi-head QNN block that replaces it at the target layer.
        The activation $\mathbf{x}$ is mixed by the constrained linear transform $\mathbf{M}_1 = \boldsymbol{\Lambda}_2\mathbf{H}\boldsymbol{\Lambda}_1$, bounded by a $\tanh$ nonlinearity, distributed across $k=32$ independent $8$-qubit QNN heads acting on disjoint $64$-dimensional subspaces, and recombined by a second constrained transform $\mathbf{M}_2$.
        The block output is scaled by $\alpha$ and returned to the residual stream.
        All remaining blocks of the base model are left untouched.}
        \label{fig:architecture}
    \end{figure*}

    The activation $\mathbf{x}$ is first mixed by a constrained linear transform $\mathbf{M}_1 = \boldsymbol{\Lambda}_2\mathbf{H}\boldsymbol{\Lambda}_1$, where $\mathbf{H}$ is the fixed normalized Sylvester--Hadamard matrix and $\boldsymbol{\Lambda}_i = \mathrm{diag}(\boldsymbol{\lambda}_i)$ are trainable diagonal matrices.
    A hyperbolic tangent then bounds the mixed features before they are embedded as rotation angles.
    Each head acts on its own subspace of the resulting vector, and a second transform $\mathbf{M}_2$ of the same form, carrying its own independent pair of diagonals, mixes the head outputs.
    The total module is thus
    \begin{equation}
        g_{\boldsymbol{\Theta}}(\mathbf{x}) = \mathbf{M}_2\,f_{\mathrm{QNNs}}\!\left(\tanh\!\left(\mathbf{M}_1 \mathbf{x}\right)\right),
    \label{eq:mhqnn}
    \end{equation}
    where $\boldsymbol{\Theta}$ collects all trainable parameters of the module. It is shown schematically in Fig.~\ref{fig:architecture}.

    Hybrid quantum--classical models risk having a classical component expressive enough to
    solve the task alone, reducing the quantum module to an
    ornament~\cite{cerezo2022challenges}; we use the constrained form above, rather
    than an unconstrained dense matrix, to guard against this. It also grows linearly
    rather than quadratically in $D$ (a three-order-of-magnitude reduction at
    $D \sim 10^3$) and the fixed Hadamard factor admits an $\mathcal{O}(D\log D)$
    evaluation via the fast Walsh--Hadamard transform. Sec.~\ref{sec:results}
    quantifies the quantum and mixer components' contributions directly.

    \subsection{Circuit design and readout}
    \label{sec:circuit}
    Individual head circuits depart from standard QNN practice in three respects, each motivated by the need to route a large hidden state through a small register.

    The first concerns how data enters the circuit.
    Contemporary QNNs rely heavily on data re-uploading, in which the same features are encoded repeatedly on the same qubits across successive layers~\cite{perezsalinas2020reuploading}; this enlarges the Fourier spectrum accessible to the model and, with it, its expressivity~\cite{schuld2021encoding}.
    We instead employ layerwise uploads~\cite{barrue2024layered}, encoding \emph{different} features on the same qubits at different layers.
    This raises the number of features a single circuit can accept without increasing the qubit count, at the cost of the additional frequencies that repetition would have supplied.
    In effect we trade accessible frequencies for a different form of nonlinearity, namely multivariate interactions among the encoded features in the expectation values of the measured Pauli strings.
    The circuits additionally apply a trainable affine map $x_i \rightarrow a_i x_i + b_i$ to each feature before encoding (here $x_i$ denotes the $i$-th component of a head's input $\tanh(\mathbf{M}_1\mathbf{x})$, not of the raw activation), so that every head can learn its own Fourier scales rather than inheriting a fixed spectrum from the encoding~\cite{jaderberg2024frequencies}.

    The second architectural deviation concerns how information leaves the circuit.
    Writing $U(\mathbf{x},\boldsymbol{\theta})$ for the full unitary of a head, the measured features are the expectation values
    \begin{equation}
        \phi_i(\mathbf{x}) = \bra{0}^{\otimes n_q} U^{\dagger}(\mathbf{x},\boldsymbol{\theta})\, P_i \, U(\mathbf{x},\boldsymbol{\theta}) \ket{0}^{\otimes n_q}, \quad i=1,\dots,m,
    \label{eq:expval}
    \end{equation}
    where $\{P_i\}_{i=1}^m$ is a pool of $m$ operators rather than the single first-order observable that is conventionally used.
    The candidate pool comprises the $4n_q$ single-qubit terms $\{I,X,Y,Z\}$ on every qubit (the $I$ terms are constant and act as a bias in the readout), together with the $9$ two-qubit strings $\{X,Y,Z\}\otimes\{X,Y,Z\}$ on each of the $n_q$ nearest-neighbour pairs of the ring (including the wraparound pair, although the entangler below is an open chain), for $13n_q$ candidates.
    It is filled to $m$ observables per head by taking all $4n_q$ single-qubit terms and drawing the remaining $m-4n_q$ from the two-qubit candidates by a fixed seeded sample without replacement; here $m = d_h = 64$, so $32$ single-qubit terms are joined by $32$ of the $72$ two-qubit candidates.
    Reading out a single string would discard an overwhelming fraction of the $4^{n_q}-1$ non-identity Pauli observables available on $n_q$ qubits, and since different strings are unequally informative for a given ansatz, committing to one in advance is a strong and largely arbitrary restriction.
    Although many of these observables commute, evaluating a pool of them provides a richer set of nonlinear basis functions for the readout than a single expectation value can.
    The two-qubit subset is drawn at random rather than ranked by an expressibility criterion because standard expressibility measures are defined over the circuit ensemble rather than over individual observables~\cite{sim2019expressibility}, leaving no efficient way to order Pauli strings for this task in advance; random selection also avoids an unexamined bias in which correlations the model can access.
    Restricting the two-qubit candidates to nearest-neighbour pairs rather than all $\binom{n_q}{2}$ keeps the pool aligned with the correlations a shallow local entangler can generate.

    The third architectural deviation concerns what is done with the measured features.
    Following common practice in QML, we place a trainable linear readout after measurement~\cite{schuld2020circuit,chen2025measure}.
    This is natural on physical grounds: observables are Hermitian, and any real linear combination of Pauli operators is Hermitian by linearity, so the readout does not take the model outside the space of valid measurements.
    The output of head $j$ is therefore
    \begin{equation}
        z^{(j)}_r = \sum_{i=1}^m W^{(j)}_{ri}\,\phi^{(j)}_i(\mathbf{x}), \quad r=1,\dots,d_h,
    \label{eq:readout}
    \end{equation}
    with $W^{(j)}$ a real $d_h \times m$ weight matrix and $\phi^{(j)}_i$ the nonlinear basis functions generated by the circuit.
    Taken together, the Pauli pool and the learned readout realize $d_h$ Hermitian observables per head, $O^{(j)}_r=\sum_i W^{(j)}_{ri}P_i$, which are optimized directly for the task.
    Because each head acts only within its own subspace, the readout across all heads is block diagonal, and its parameter count is small relative to that of a dense $D \times D$ readout.

\begin{figure}
        \centering
        \includegraphics[width=\linewidth]{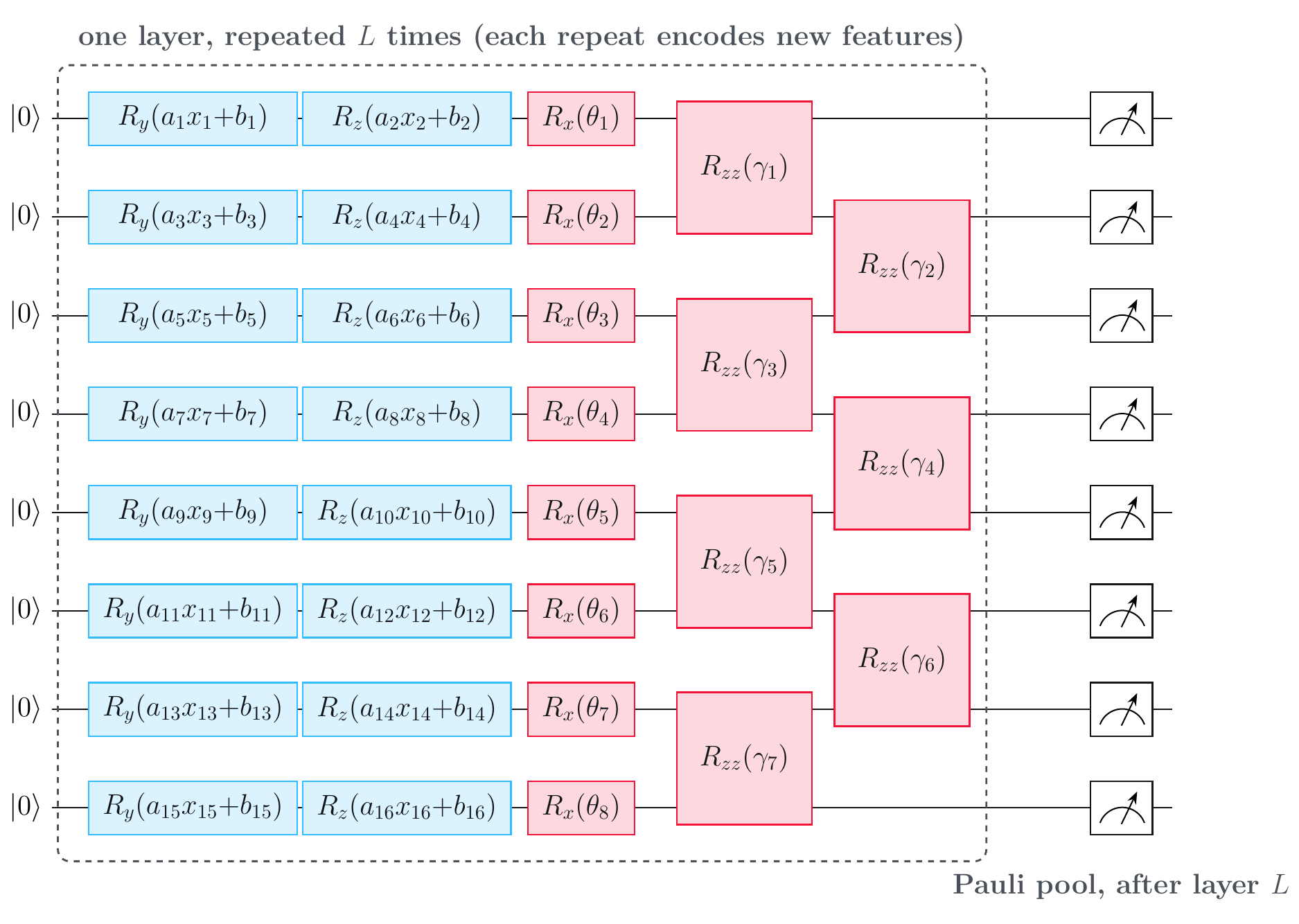}
        \caption{%
        One layer of an eight-qubit head circuit of the multi-head QNN.
        Each qubit receives two features through dense angle encoding, as $R_y$ and $R_z$ rotations whose angles are trainable affine functions $a_ix_i+b_i$ of the input, allowing each head to learn its own Fourier scales~\cite{jaderberg2024frequencies}.
        A layer of variational $R_x(\theta)$ gates follows, and entanglement is generated by an open chain of $n_q-1=7$ separate nearest-neighbour $R_{zz}(\gamma_j)$ gates with no wraparound edge, matching the linear-connectivity ansatz described in the text and the two-qubit gate counts reported in Sec.~\ref{sec:hardware} (the gates commute, so the chain is drawn in two parallel steps).
        The dashed box is repeated $L=4$ times in the deployed model; successive layers encode distinct features rather than re-uploading the same ones, so that a single circuit ingests $2n_qL = 64$ features, and the Pauli observables are measured once, after layer $L$.}
        \label{fig:circuit}
    \end{figure}

    For the ansatz itself we use a DAE circuit in which each layer encodes two features per qubit through $R_y$ and $R_z$ rotations, applies a layer of variational $R_x$ gates, and entangles with an open, linearly connected chain of $n_q-1$ $R_{zz}$ gates (no wraparound edge).
    We adopt this construction for its hardware efficiency, its simplicity, and its avoidance of costly routing: all two-body gates act on nearest neighbours, so no SWAP networks are required on a linear-connectivity device~\cite{kandala2017hardware,leone2024practical}.
    This choice is made for the device rather than justified after the fact, and Sec.~\ref{sec:hardware} reports what it buys: with $R_{zz}$ made native by the processor's fractional gate set, the head's logical two-qubit rotations map one-to-one onto device gates with zero routing overhead.
    A single layer of the eight-qubit circuit is shown in Fig.~\ref{fig:circuit}.

    \subsection{Instantiation in Qwen3-30B-A3B}

    In this paper, we use multi-head QNNs to replace transformer blocks of the open-weight model Qwen3-30B-A3B, a $30$-billion-parameter mixture-of-experts transformer in which only the top-ranked experts are activated for any given token, so that roughly $3$ billion parameters are active per token~\cite{jacobs1991moe,shazeer2017moe,fedus2022switch}.
    As illustrated in Fig.~\ref{fig:architecture}, we replace a target transformer block with a multi-head QNN block whose contribution to the residual stream is scaled by a learned factor $\alpha$~\cite{bachlechner2021rezero,touvron2021going},
    \begin{equation}
        \mathbf{y} = \mathbf{x} + \alpha\, g_{\boldsymbol{\Theta}}(\mathbf{x}),
    \label{eq:mhqnn_install}
    \end{equation}
    where $\mathbf{y}$ is the transformed activation.
    The role of $\alpha$ is to stabilize training and to keep the residual addition at a scale the surrounding network can absorb.
    In practice $\alpha$ settles to small values over training, and its magnitude is \emph{not} on its own a valid measure of the block's influence, because the diagonal factors of $\mathbf{M}_1$ and $\mathbf{M}_2$ are free to rescale the signal internally; Sec.~\ref{sec:discussion} sets out the one limit in which the gate is nonetheless informative.

    In practice we set $n_q=8$, $L=4$ and $k=32$, so that $D = 2n_qLk = 2048$ matches the hidden dimension of Qwen3-30B-A3B and each head is responsible for a $d_h = 2n_qL = 64$-dimensional subspace of the activations.
    Table~\ref{tab:params} summarizes the resulting parameter budget; Appendix~\ref{annex:model} gives the same breakdown for every ablation condition and for the register-width and upload-depth variants, together with the fixed min--max interface buffers that Eq.~\eqref{eq:mhqnn_install} suppresses.

    The entire block carries $1.45\times10^{5}$ trainable parameters, of which the learned readout accounts for roughly $90\%$ at this choice of $m$; the two constrained mixers together contribute $8\,192$, against the $8.4\times10^{6}$ that two unconstrained $2048\times2048$ matrices would require, and against the $6.23\times10^{8}$ of the block being replaced.
    Because the readout scales as $d_h \cdot m \cdot k$ while the circuit parameters scale as $\mathcal{O}(n_qLk)$, the readout dominates the budget whenever $m$ is comparable to $d_h$, and the parameter accounting below should be read as specific to $m=d_h=64$ rather than as a property of the architecture.

    We probe three levels of replacement.
    The first is single-block replacement, in which one target transformer block is replaced by a single multi-head QNN block as in Eq.~\eqref{eq:mhqnn_install}, isolating the contribution of one instance of the module at one depth.
    The second replacement level is \emph{block fusion}, in which a contiguous span of several transformer blocks is collapsed into a \emph{single} multi-head QNN block trained to reproduce the joint effect of the span.
    The third is disjoint multi-block replacement, in which several non-adjacent target blocks are each replaced by their own independent multi-head QNN block within the same model.
    Together these scenarios probe three separate questions: whether a single block is replaceable, whether multiple adjacent blocks are jointly replaceable, and whether replacement composes across depth.

    Training proceeds in two stages. The base model's weights are frozen throughout, and
    the multi-head QNN is the only trainable component.

    In Stage 1 we train the multi-head QNN to reproduce the output of the
    target transformer block, without regard to the output of the full model. We pass
    calibration data through the network and record the activation entering the target
    block as $\mathbf{x}$, and the activation leaving it (that is, entering the
    following block) as $\mathbf{y}$. Writing $\tilde{\mathbf{y}}$ for the
    multi-head QNN's output on the same input $\mathbf{x}$, we minimize the combination of a mean squared error (MSE) and a cosine-similarity term,
    \begin{equation}
    \mathcal{L} = \mathrm{MSE}(\mathbf{y}, \tilde{\mathbf{y}})
                - 0.1 \, \cos(\mathbf{y}, \tilde{\mathbf{y}}),
    \end{equation}
    where the MSE enforces agreement in magnitude and the cosine term,
    weighted by $0.1$, enforces agreement in direction.

    In Stage 2 we fine-tune the resulting module end to end on the model's usual
    next-token prediction loss. The multi-head QNN remains the only trainable component;
    all other transformer blocks and the MoE routers stay frozen.

    Both stages use \emph{webmix}, an internal training mixture of web, encyclopedic and
    educational text sources (Wikipedia, FineWeb-Edu, DCLM-Edu and smaller math and
    synthetic-text components), and the Adam optimizer~\cite{kingma2015adam}. The per-stage optimizer settings, including learning rate, batch size and step budget for each stage, are given in Appendix~\ref{annex:protocol}.
    Checkpoints are selected by validation perplexity on a disjoint $256$-sequence tail evaluated every $500$ steps; the full protocol and the separation between training, validation and evaluation corpora are set out in Appendix~\ref{annex:protocol}. Since webmix contains Wikipedia, we checked the overlap: no $13$-gram of WikiText-2 test and fewer than $0.005\%$ of those of C4 validation occur in the training subset (Appendix~\ref{annex:protocol}).
    Head circuits are simulated with a tensor-contraction-based statevector simulator implemented in PyTorch~\cite{paszke2019pytorch}, allowing training to proceed entirely on classical hardware.
    Hardware inference, in which the trained model is deployed with the head circuits executed on a quantum processor by packing the $k$ heads onto disjoint patches of the device, is reported in Sec.~\ref{sec:hardware}.

    \begin{table}
    \caption{%
    Trainable parameter budget of a single multi-head QNN block at $n_q=8$, $L=4$ and $k=32$, corresponding to a hidden dimension of $D = 2n_qLk = 2048$.
    Frozen buffers, including the fixed Sylvester--Hadamard factor $\mathbf{H}$, are excluded.
    The learned readout has shape $d_h \times m$ per head, with $m=2n_qL=64$ in this instantiation (Sec.~\ref{sec:circuit}); in general the readout parameter count is $d_h \cdot m \cdot k$.
    The two constrained mixers together account for $8\,192$ parameters, compared with $8.4\times10^{6}$ for two unconstrained dense matrices of the same shape.}
    \label{tab:params}
    \begin{ruledtabular}
    \begin{tabular}{llr}
    Component & Expression & Count \\
    \colrule
    Input mixer $\mathbf{M}_1$      & $2D$              & $4\,096$ \\
    Encoding scales and biases      & $4n_qLk$          & $4\,096$ \\
    Variational $R_x$ gates         & $n_qLk$           & $1\,024$ \\
    $R_{zz}$ entanglers             & $(n_q-1)Lk$       & $896$ \\
    Learned readout                 & $d_h \cdot m \cdot k$ & $131\,072$ \\
    Output mixer $\mathbf{M}_2$     & $2D$              & $4\,096$ \\
    Residual scale $\alpha$         & $1$               & $1$ \\
    \colrule
    Total                           &                   & $145\,281$ \\
    \end{tabular}
    \end{ruledtabular}
    \end{table}

    \section{Results}
    \label{sec:results}

    We measure generalization with token-level perplexity (PPL) on two corpora held out from
    training: the WikiText-2-raw test set~\cite{merity2017wikitext} and the C4 English
    validation set~\cite{raffel2020c4}. We also report perplexity on
    \texttt{webmix-recon}, a $1024$-sequence sample drawn from the training file itself.
    This shows how closely the multi-head QNN replacement fits the data it was trained on but is not
    evidence of generalization. We test
    downstream ability on seven standard tasks with the
    \texttt{lm-evaluation-harness}~\cite{gao2023lmeval} (\texttt{lm-eval}): LAMBADA~\cite{paperno2016lambada},
    HellaSwag~\cite{zellers2019hellaswag}, ARC-easy and ARC-challenge~\cite{clark2018arc},
    PIQA~\cite{bisk2020piqa}, WinoGrande~\cite{sakaguchi2020winogrande}, and
    BoolQ~\cite{clark2019boolq}. Finally, we measure how closely each replacement tracks
    the original model, using top-1 next-token agreement on C4. 

    Three points govern how the results below should be read. First, each of these measurements carries two kinds of noise, and we keep them separate. Evaluation
    noise comes from the choice of text: scoring the same trained model on a different
    sample gives a slightly different result. We quantify it by rescoring the model on many
    resampled versions of the evaluation text and reporting the range that covers $95\%$ of
    the resulting scores. Where per-document scores are available for both models in a
    comparison, we evaluate them on the same documents, which cancels out differences in
    document difficulty; otherwise we fall back on a worst-case bound, which is wide and
    registers only large differences (Appendix~\ref{annex:resolution}). Training noise comes
    from the random seed: the same configuration trained twice gives slightly different
    models. We quantify it as the standard deviation over three seeds. We never merge the
    two into a single figure. A value quoted as a mean with a standard deviation is taken
    over the three seeds, individual-seed values are given in the appendices, and we call a
    comparison \emph{resolved} when its $95\%$ interval excludes zero. Full estimators are
    given in Appendix~\ref{annex:protocol}, Sec.~\ref{annex:uncertainty}.

    Second, there is a training asymmetry: the original model is not fine-tuned on the data used to train the replacement module. Specifically, each replacement
    module is trained for $20\,000$ steps on webmix, while the original model receives no
    further training, so gains at small span sizes may partly reflect this extra training
    rather than the module itself. The \emph{identity-skip} baseline (the same layers deleted,
    with nothing trained in their place) does not isolate this effect, since nothing is trained; it marks what removal alone costs, and we quote it wherever
    the comparison is at issue. A fine-tuned base model or a healed truncated model would be the matched controls, and neither is part of this study.

    Third, different experiments target different depths. Block fusion replaces layers in
    the redundant middle of the network, whereas single-block and disjoint replacement
    target more sensitive layers, early layers 1--6 and 8 and the tail layers 38--46. These are different regimes,
    and we report them separately rather than comparing them directly.

    Target layers are chosen by identity-skip: each of the $48$ decoder layers is turned into the identity map in turn, by zeroing its additive contribution to the residual stream, and WikiText-2 perplexity is re-measured, the model being restored between measurements. We take the resulting $\Delta\mathrm{PPL}$ against the unmodified baseline of $8.810$ as the necessity signal; this ranking uses \texttt{lm-eval} word-level perplexity, whereas the results reported below use token-level perplexity (the one exception being the noise check of Sec.~\ref{sec:hardware}), for which the unmodified baseline is $8.672$. The ranking is measured on the WikiText-2 test split, which is also one of the evaluation sets below; every comparison is made at the same layers across conditions, so the choice of layers does not favour any condition, but it was not made on held-out data, and selecting the span on WikiText-2 can favour every replaced model over the unmodified one on that set; C4 is the selection-independent check. Repeating the ranking on the $256$-sequence validation tail of the training corpus (Appendix~\ref{annex:protocol}), disjoint from every evaluation set, gives nearly the same order: the Spearman rank correlation with the WikiText-2 ranking over the $46$ interior layers is $0.95$, layer $1$ is again the most necessary ($+2.07$ on a baseline of $6.89$), the tail layers $42$ to $46$ and layer $3$ follow, and layers $10$ to $17$ rank $30$th to $46$th. The sets of the eight and sixteen most necessary layers differ from the WikiText-2 ones by one layer each. The two boundary layers dominate and are excluded as replacement targets ($\Delta\mathrm{PPL} = +84.97$ at layer $0$ and $+11.55$ at layer $47$); among the remainder, layer $1$ is the most necessary ($+5.06$), followed by the tail layers $45$, $44$, $43$, $46$ and $42$ ($+1.61$ down to $+0.84$), while layers $10$ to $17$ are among the most redundant ($+0.03$ to $+0.14$). Single-block and disjoint multi-block replacement accordingly target the head of this ranking, the $N$-block sets being its first $N$ entries; block fusion targets the contiguous span beginning at layer $10$, in the redundant middle.

    Next, subsections A--C below establish that the multi-head QNN substitution works, how far it can
    be pushed, and where it stops working, and subsection D describes model execution on an actual quantum device.

    \subsection{Block fusion (main result)}
    \label{sec:fusion}

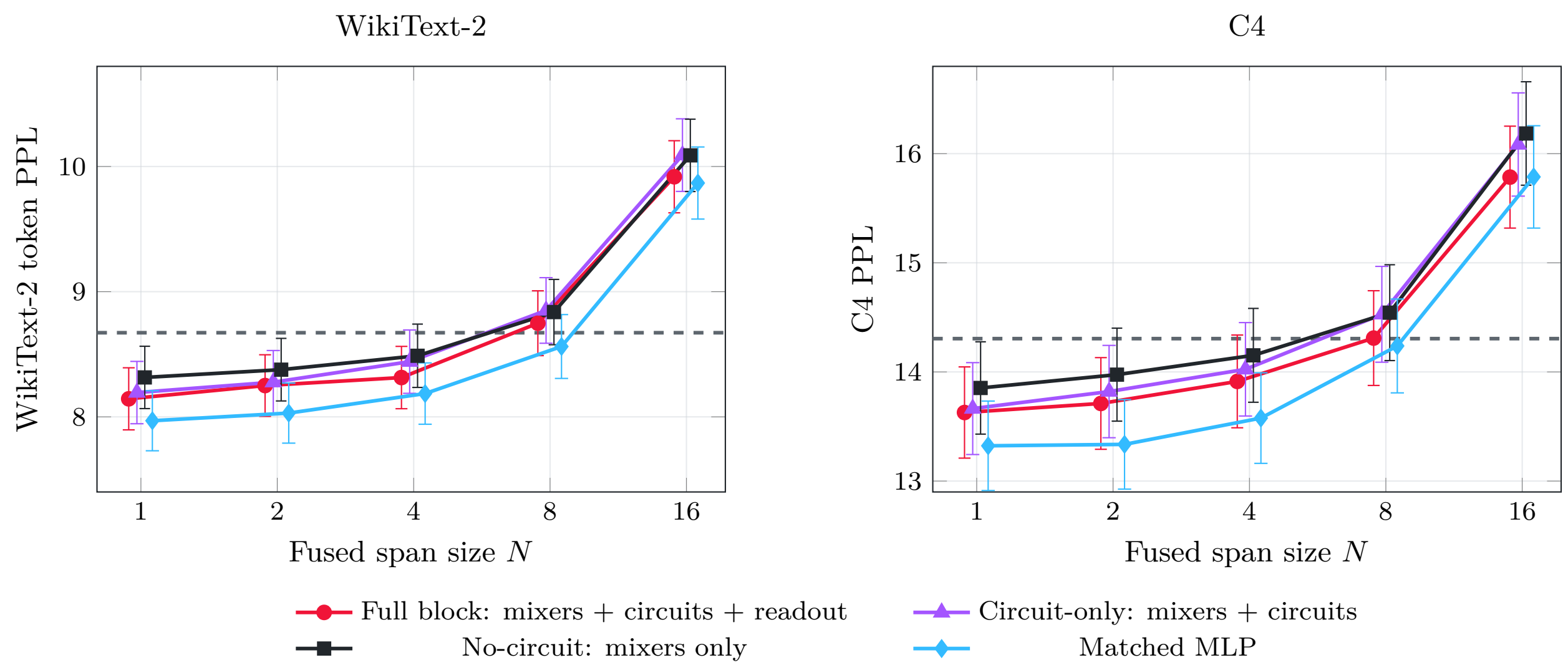
\begin{figure*}[t]
        \centering
        \begin{tikzpicture}
        \begin{groupplot}[
            group style={
                group size=2 by 1,
                horizontal sep=2.2cm,
            },
            width=0.46\textwidth,
            height=0.34\textwidth,
            xmode=log,
            xtick={1,2,4,8,16},
            xticklabels={1,2,4,8,16},
            xmin=0.80, xmax=19.5,
            xlabel={Fused span size $N$},
            grid=major,
            grid style={mvcgray500, opacity=0.5},
            tick label style={font=\footnotesize},
            label style={font=\small},
            axis line style={mvcgray900},
            every axis plot/.append style={line width=1.1pt},
            error bars/y dir=both,
            error bars/y explicit,
        ]

        \nextgroupplot[
            ylabel={WikiText-2 token PPL},
            ymin=7.4, ymax=10.8,
            title={\small WikiText-2},
        ]

        \addplot[mvcgray700, dashed, no marks, forget plot]
            coordinates {(0.80,8.672) (19.5,8.672)};

        \addplot[mvcrose, mark=*, mark size=1.8pt]
            plot[error bars/.cd, y dir=both, y explicit]
            coordinates {
            (0.94,8.144) +- (0,0.248)
            (1.88,8.250) +- (0,0.246)
            (3.76,8.314) +- (0,0.249)
            (7.52,8.748) +- (0,0.259)
            (15.04,9.918) +- (0,0.288)
        };

        \addplot[mvcpurple, mark=triangle*, mark size=2pt]
            plot[error bars/.cd, y dir=both, y explicit]
            coordinates {
            (0.98,8.194) +- (0,0.249)
            (1.96,8.277) +- (0,0.253)
            (3.92,8.441) +- (0,0.253)
            (7.84,8.850) +- (0,0.262)
            (15.68,10.091) +- (0,0.290)
        };

        \addplot[mvcgray900, mark=square*, mark size=1.6pt]
            plot[error bars/.cd, y dir=both, y explicit]
            coordinates {
            (1.02,8.315) +- (0,0.249)
            (2.04,8.377) +- (0,0.250)
            (4.08,8.488) +- (0,0.253)
            (8.16,8.837) +- (0,0.261)
            (16.32,10.089) +- (0,0.289)
        };

        \addplot[mvcblue, mark=diamond*, mark size=2pt]
            plot[error bars/.cd, y dir=both, y explicit]
            coordinates {
            (1.06,7.969) +- (0,0.240)
            (2.12,8.030) +- (0,0.240)
            (4.24,8.186) +- (0,0.245)
            (8.48,8.562) +- (0,0.255)
            (16.96,9.868) +- (0,0.288)
        };

        \nextgroupplot[
            ylabel={C4 PPL},
            ymin=12.9, ymax=16.8,
            title={\small C4},
            legend to name={fusionlegend},
            legend style={
                legend columns=2,
                font=\footnotesize,
                draw=none,
                /tikz/every even column/.append style={column sep=0.6cm},
            },
        ]

        \addplot[mvcgray700, dashed, no marks, forget plot]
            coordinates {(0.80,14.305) (19.5,14.305)};

        \addplot[mvcrose, mark=*, mark size=1.8pt]
            plot[error bars/.cd, y dir=both, y explicit]
            coordinates {
            (0.94,13.628) +- (0,0.418)
            (1.88,13.711) +- (0,0.420)
            (3.76,13.913) +- (0,0.425)
            (7.52,14.310) +- (0,0.434)
            (15.04,15.785) +- (0,0.467)
        };

        \addplot[mvcpurple, mark=triangle*, mark size=2pt]
            plot[error bars/.cd, y dir=both, y explicit]
            coordinates {
            (0.98,13.664) +- (0,0.421)
            (1.96,13.820) +- (0,0.423)
            (3.92,14.024) +- (0,0.428)
            (7.84,14.528) +- (0,0.439)
            (15.68,16.084) +- (0,0.473)
        };

        \addplot[mvcgray900, mark=square*, mark size=1.6pt]
            plot[error bars/.cd, y dir=both, y explicit]
            coordinates {
            (1.02,13.853) +- (0,0.423)
            (2.04,13.975) +- (0,0.426)
            (4.08,14.152) +- (0,0.430)
            (8.16,14.543) +- (0,0.439)
            (16.32,16.185) +- (0,0.474)
        };

        \addplot[mvcblue, mark=diamond*, mark size=2pt]
            plot[error bars/.cd, y dir=both, y explicit]
            coordinates {
            (1.06,13.323) +- (0,0.410)
            (2.12,13.336) +- (0,0.410)
            (4.24,13.577) +- (0,0.415)
            (8.48,14.238) +- (0,0.430)
            (16.96,15.787) +- (0,0.469)
        };

        \legend{
            Full block: mixers $+$ circuits $+$ readout,
            Circuit-only: mixers $+$ circuits,
            No-circuit: mixers only,
            Matched MLP
        }

        \end{groupplot}

        \node[anchor=north] at ($(group c1r1.south)!0.5!(group c2r1.south)$) [yshift=-0.9cm] {\NoHyper\ref{fusionlegend}\endNoHyper};

        \end{tikzpicture}

        \caption{%
        Test perplexity as a function of fused span size $N$ on WikiText-2 and C4, for a single multi-head QNN block of $145\,281$ parameters, fixed across $N$, trained to replace a contiguous span of $N$ transformer blocks under the four matched conditions defined in the text, all sharing an identical schedule and seed.
        Series are horizontally offset at each $N$ for legibility; error bars show $95\%$ bootstrap confidence intervals, and dashed lines mark the unmodified base model.
        }
        \label{fig:fusion_ablation}
    \end{figure*}

    Our principal result is that one multi-head QNN block can replace a contiguous span of several transformer blocks at once, and that the cost of doing so grows slowly with the size of the span.
    A multi-head QNN of fixed size ($145\,281$ trainable parameters) is trained to replace the contiguous span from transformer block 10 to block $(9{+}N)$ for $N \in \{1,2,4,8,16\}$.
    Test perplexity appears as the red series in Fig.~\ref{fig:fusion_ablation}.

    Measured against the identity-skip baseline (simply deleting the same span of blocks), the multi-head QNN performs better on all three evaluation sets at all five spans $N \in \{1,2,4,8,16\}$.
    The margin is smaller at $N=1$, where WikiText-2 perplexity falls from $8.791$ to $8.144$, but grows at $N=16$, where it falls from $13.115$ to $9.918$. 
    Degradation with span size is gradual rather than abrupt: WikiText-2 perplexity rises from $8.144$ to $8.748$ between $N=1$ and $N=8$, while top-1 agreement with the base model falls from $0.883$ to $0.838$ over the same interval.

    The resulting hybrid model is even better than the original model at $N=1$, $2$ and $4$ on both test corpora, reaches the crossover at $N=8$, where it is worse on WikiText-2 by $0.9\%$ and unresolved on C4 and on \texttt{webmix-recon}, and is resolved worse at $N=16$. The gains at $N \leq 4$ should be read with the training asymmetry in mind, and at $N=8$ we claim only near-parity on the held-out sets.
    Repeating $N=4$ and $N=8$ at a second training seed reproduces test log-likelihood to within $0.0016$, so the shape of the curve in Fig.~\ref{fig:fusion_ablation} is not a seed artifact.

    \subsubsection*{Compression}
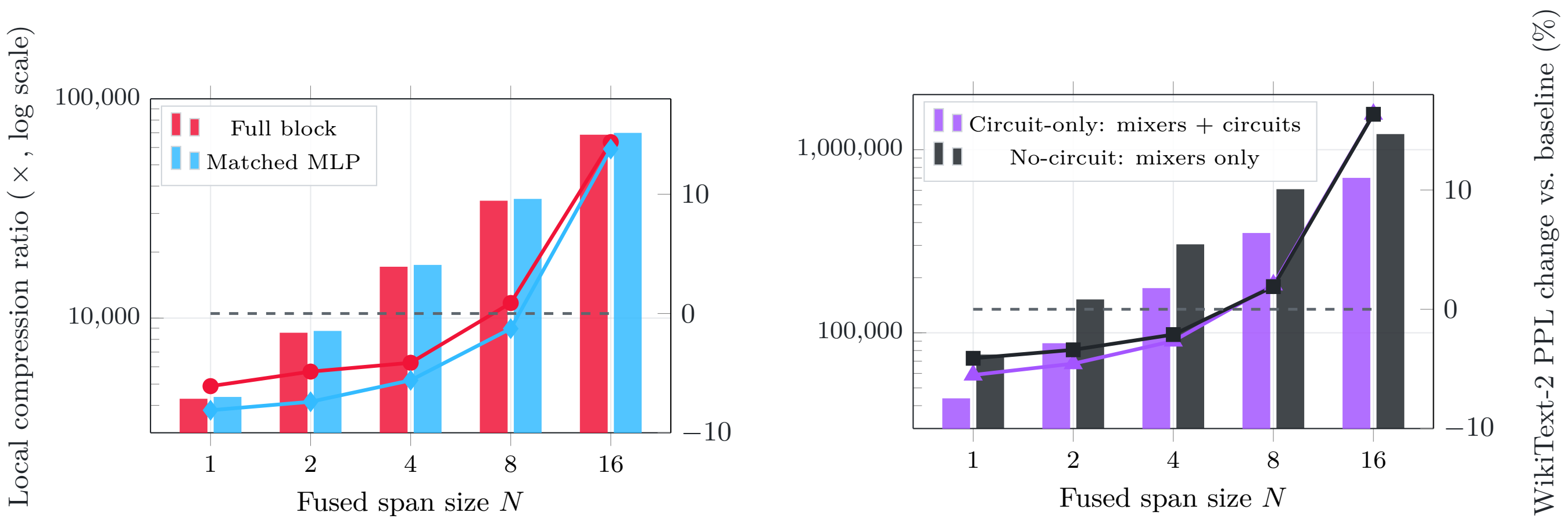
\begin{figure*}[t]
        \centering

        \begin{minipage}[b]{0.455\textwidth}
        \centering
        \hspace*{-0.1\textwidth}%
        \begin{tikzpicture}

        \begin{axis}[
            width=0.92\linewidth,
            height=0.66\linewidth,
            ybar,
            bar width=9pt,
            xlabel={Fused span size $N$},
            ylabel={Local compression ratio ($\times$, log scale)},
            ymode=log,
            log ticks with fixed point,
            symbolic x coords={1,2,4,8,16},
            xtick=data,
            ymin=3e3, ymax=1e5,
            axis y line*=left,
            ylabel style={mvcgray900},
            yticklabel style={mvcgray900},
            grid=major,
            grid style={mvcgray500, opacity=0.5},
            tick label style={font=\footnotesize},
            label style={font=\small},
            axis line style={mvcgray900},
            enlarge x limits=0.15,
            legend style={
                at={(0.02,0.98)},
                anchor=north west,
                font=\scriptsize,
                draw=mvcgray500,
                fill=white,
                fill opacity=0.9,
                text opacity=1,
            },
            legend columns=1,
        ]

        \addplot[draw=none, fill=mvcrose, fill opacity=0.85]
            coordinates {
                (1,4288.5)
                (2,8577.0)
                (4,17154.0)
                (8,34308.0)
                (16,68616.0)
            };

        \addplot[draw=none, fill=mvcblue, fill opacity=0.85]
            coordinates {
                (1,4371.0)
                (2,8742.1)
                (4,17484.1)
                (8,34968.2)
                (16,69936.5)
            };

        \legend{
            Full block,
            Matched MLP
        }
        \end{axis}

        \begin{axis}[
            width=0.92\linewidth,
            height=0.66\linewidth,
            symbolic x coords={1,2,4,8,16},
            xtick=data,
            xticklabels={,,},
            ymin=-10, ymax=18,
            axis y line*=right,
            axis x line=none,
            yticklabel style={mvcgray900},
            enlarge x limits=0.15,
        ]

        \addplot[
            mark=*,
            mark size=2pt,
            mvcrose,
            line width=1.2pt
        ]
            coordinates {
                (1,-6.09)
                (2,-4.87)
                (4,-4.13)
                (8,0.88)
                (16,14.37)
            };

        \addplot[
            mark=diamond*,
            mark size=2.4pt,
            mvcblue,
            line width=1.2pt
        ]
            coordinates {
                (1,-8.11)
                (2,-7.40)
                (4,-5.60)
                (8,-1.27)
                (16,13.79)
            };

        \draw[mvcgray700, dashed, line width=0.9pt]
            (axis cs:1,0) -- (axis cs:16,0);

        \end{axis}

        \end{tikzpicture}
        \end{minipage}%
        \hspace{0.015\textwidth}%
        \begin{minipage}[b]{0.455\textwidth}
        \centering
        \begin{tikzpicture}

        \begin{axis}[
            width=0.92\linewidth,
            height=0.66\linewidth,
            ybar,
            bar width=9pt,
            xlabel={Fused span size $N$},
            ymode=log,
            log ticks with fixed point,
            symbolic x coords={1,2,4,8,16},
            xtick=data,
            ymin=3e4, ymax=2e6,
            axis y line*=left,
            yticklabel style={mvcgray900},
            grid=major,
            grid style={mvcgray500, opacity=0.5},
            tick label style={font=\footnotesize},
            label style={font=\small},
            axis line style={mvcgray900},
            enlarge x limits=0.15,
            legend style={
                at={(0.02,0.98)},
                anchor=north west,
                font=\scriptsize,
                draw=mvcgray500,
                fill=white,
                fill opacity=0.9,
                text opacity=1,
            },
            legend columns=1,
        ]

        \addplot[draw=none, fill=mvcpurple, fill opacity=0.85]
            coordinates {
                (1,43855.7)
                (2,87711.4)
                (4,175422.8)
                (8,350845.6)
                (16,701691.3)
            };

        \addplot[draw=none, fill=mvcgray900, fill opacity=0.85]
            coordinates {
                (1,76054.7)
                (2,152109.3)
                (4,304218.7)
                (8,608437.3)
                (16,1216874.7)
            };

        \legend{
            Circuit-only: mixers $+$ circuits,
            No-circuit: mixers only
        }
        \end{axis}

        \begin{axis}[
            width=0.92\linewidth,
            height=0.66\linewidth,
            symbolic x coords={1,2,4,8,16},
            xtick=data,
            xticklabels={,,},
            ymin=-10, ymax=18,
            axis y line*=right,
            axis x line=none,
            ylabel={WikiText-2 PPL change vs.\ baseline (\%)},
            ylabel style={mvcgray900, yshift=-6pt},
            yticklabel style={mvcgray900},
            enlarge x limits=0.15,
        ]

        \addplot[
            mark=triangle*,
            mark size=2.4pt,
            mvcpurple,
            line width=1.2pt
        ]
            coordinates {
                (1,-5.51)
                (2,-4.56)
                (4,-2.66)
                (8,2.05)
                (16,16.36)
            };

        \addplot[
            mark=square*,
            mark size=1.8pt,
            mvcgray900,
            line width=1.2pt
        ]
            coordinates {
                (1,-4.11)
                (2,-3.40)
                (4,-2.12)
                (8,1.90)
                (16,16.35)
            };

        \draw[mvcgray700, dashed, line width=0.9pt]
            (axis cs:1,0) -- (axis cs:16,0);

        \end{axis}

        \end{tikzpicture}
        \end{minipage}

        \caption{%
        Local compression ratio (bars, left axis, logarithmic) and WikiText-2 token perplexity change relative to the base model (lines, right axis) against fused span size $N$.
        Compression ratio is the parameter count of the $N$ removed transformer blocks (total-parameter basis), $N \times 623\,120\,640$, divided by the trainable replacement count: $145\,281$ for the full block, $142\,537$ for the MLP, $14\,209$ for circuit-only and $8\,193$ for the no-circuit control.
        \textbf{Left:} the full block and the parameter-matched MLP occupy nearly identical compression ratios, the MLP being the lower in perplexity.
        \textbf{Right:} the readout-free conditions reach substantially higher compression ratios at a correspondingly higher perplexity cost, the no-circuit control achieving the highest of any condition tested on both counts.
        Dashed gray lines mark parity with the unmodified model; the two panels use different compression-ratio ranges for the two parameter regimes.
        }
        \label{fig:compression}

    \end{figure*}

    Figure~\ref{fig:compression}(left) shows the local compression ratio.
    Each removed transformer block contains $6.23\times10^{8}$ parameters while the multi-head QNN carries $1.45\times10^{5}$ parameters (regardless of span size), so the compression ratio is set entirely by how many layers the span covers.
    At $N=8$ this removes $16.33\%$ of the full $30.5$-billion-parameter model for a reduction factor of $1.195$, and at $N=16$ it removes $32.65\%$ for a factor of $1.485$; on the active-parameter basis, which governs inference cost, the same spans remove $13.57\%$ and $27.14\%$ of the $3.35$ billion active parameters. The local ratio of Fig.~\ref{fig:compression} is set by the size of the module, and identity-skip removes the same $16.33\%$ at $N=8$; Appendix~\ref{annex:compression} gives the accounting.
    $N=8$ is the widest span at which the replaced model stays within about $1\%$ of the
    original in perplexity on both held-out sets (resolved worse on WikiText-2, unresolved on C4), and $N=4$ is the widest at which neither held-out set is resolved worse; we present $N=16$ only as the outer edge
    of graceful degradation. Fig.~\ref{fig:compression}(left) makes the gap concrete: WikiText-2
    perplexity is essentially at parity at $N=8$ ($+0.9\%$) and far above it at $N=16$
    ($+14.4\%$).

    \subsubsection*{Additive contribution of the quantum circuits}

    The results above establish that the multi-head QNN as a whole
    recovers the function of the displaced blocks.
    We now ask how much of that recovery each part contributes, comparing the full block
    with variants obtained by deleting named components of
    Fig.~\ref{fig:architecture} and training everything else identically.

    The \emph{circuit-only} variant deletes the learned readout and nothing else: the
    $m=64$ Pauli expectation values of each of the $32$ heads pass straight to the output
    mixer $\mathbf{M}_2$, leaving $14\,209$ parameters.
    The \emph{no-circuit} variant deletes the heads and the readout as well, reducing the
    block to its classical path
    $\mathbf{y} = \mathbf{x} + \alpha\,\mathbf{M}_2\tanh(\mathbf{M}_1\mathbf{x})$
    at $8\,193$ parameters, and isolates what the heads and readout add on top of the
    mixing alone.
    As an external classical reference, we also train a parameter-matched MLP.

    The full block beats the no-circuit variant at every span size on every evaluation set, as a single-seed point estimate.
    The no-circuit variant nonetheless improves on the base model at $N \leq 4$, reaching
    $8.315$ on WikiText-2 at $N=1$ against $8.672$, and stays below identity-skip at every span size. However, when replacing a single highly sensitive layer (a different layer from the $N=1$ case; see Sec.~\ref{sec:single}), the same circuit-free replacement ends up above
    identity-skip on test WikiText-2 at all three seeds (Table~\ref{tab:annex_single_seeds}), so we read the improvement
    at $N \leq 4$ with caution, as reflecting the low sensitivity of the blocks replaced
    and the replacement's fine-tuning on a tailored dataset rather than any capability of
    the mixers themselves.

    The circuit-only results are more consequential: at $N=4$ the variant reaches $8.441$
    on WikiText-2 and $14.024$ on C4, within $2\%$ of the full block, while carrying
    $14\,209$ trainable parameters against the $2.49\times10^{9}$ of the span it replaces,
    a local compression ratio of $1.75\times10^{5}$
    (Fig.~\ref{fig:compression}(right)); the learned readout, $90\%$ of the full
    block's parameters, purchases a further improvement that is consistent in sign but
    modest.
    Isolating the circuit itself, the circuit-only variant leads the no-circuit variant, as a single-seed point estimate, at
    $N=1$, $2$ and $4$ on all three evaluation sets and the two are statistically
    indistinguishable beyond $N=4$, the circuit's isolated contribution being unresolved at
    these depths.
    The matched MLP leads the full block in perplexity in all but one comparison (C4 at $N=16$, where the two differ by $0.002$); however,
    downstream accuracy and top-1 agreement do not consistently show this lead (Appendix~\ref{annex:fusion}); see Sec.~\ref{sec:discussion} for further discussion.

    Complete per-condition values, including a frozen-readout arm omitted from Fig.~\ref{fig:fusion_ablation} for legibility, are given in Appendix~\ref{annex:fusion}.

    The span ablation is consistent with a contribution from the heads but, being single-seed and bound-only, cannot resolve it, and the displaced blocks were not highly sensitive. The next subsection resolves it by replacing a single, highly sensitive block.

    \subsection{Single-block replacement}
    \label{sec:single}

    The previous experiment could not resolve the circuit's contribution because the
    displaced blocks had relatively little effect on model performance. When the blocks
    matter little, even a poor replacement looks adequate, and the still smaller
    contribution of the circuit within that replacement is harder to resolve. We therefore repeat the experiment on a highly sensitive block, layer 1 (layers are indexed from $0$), training three seeds per condition. At seed 42 we retain per-document scores for
    every condition, so the conditions can be compared on the same documents. This resolves
    the comparison that the fusion ablation could not.
    Perplexity and top-1 agreement are given in Table~\ref{tab:single_block_ppl} and
    downstream accuracy in Table~\ref{tab:single_block_downstream}; the individual
    runs behind both appear in Appendix~\ref{annex:single}.

    Removing the quantum heads while holding the mixing scaffold, the interface, the residual gate and the training schedule fixed is costly for this block replacement.
    The full multi-head QNN reaches $9.309 \pm 0.079$ on WikiText-2 against $12.787 \pm 0.838$ for the no-circuit control, a mean separation of $3.5$ perplexity points that ranges from $2.8$ to $4.4$ across the individual seeds, resolved on all nine combinations of seed and evaluation set.
    The full multi-head QNN is likewise resolved better than identity-skip ($11.941$) on all three evaluation sets.
    This is the cleanest evidence in the paper that the quantum heads perform work beyond that of the classical components; Sec.~\ref{sec:disc_ablations} shows that this does not require the circuits to be trained or entangled. The no-circuit control removes the circuits and the readout together; in the readout ablations of Table~\ref{tab:annex_readout} (seed 42), removing only the readout raises WikiText-2 perplexity from $9.230$ to $9.508$ (unresolved), and also removing the circuits raises it to $12.050$.
    The no-circuit variant is also far less stable across seeds than the full block, at $\pm0.838$ against $\pm0.079$ on WikiText-2 perplexity; this dispersion is discussed in Sec.~\ref{sec:discussion}.

    In the opposite direction the matched MLP keeps its advantage, beating the full multi-head QNN by $2.9$ to $6.7\%$ in perplexity at all nine combinations of seed and evaluation set, although only seed 42 is resolved. Both conditions train stably, and the gap between them exceeds both the spread across seeds and the within-run evaluation noise.

    Downstream, the full multi-head QNN and the matched MLP agree within seed dispersion on six of the seven tasks (LAMBADA is the exception), but how much they recover of what identity-skip forfeits depends on the task.
    On WinoGrande both recover most of the gap, reaching $0.715$ and $0.711$ against the identity value of $0.663$ and a base value of $0.728$ ($80\%$ and $74\%$ of the gap); on LAMBADA, the task most dependent on long-range context, the recovery is partial, at $0.668$ and $0.684$ against $0.653$ and a base value of $0.711$ ($26\%$ and $53\%$), a contrast we return to in Sec.~\ref{sec:discussion}.
    BoolQ behaves differently: the full block and the matched MLP land at $0.833$ and $0.830$ against the base model's $0.885$, while identity-skip at $0.881$ and the near-identity no-circuit variant at $0.877$ barely fall, which is consistent with fine-tuning on webmix rather than the architecture.

    \begin{table*}[t]
    \caption{%
    Single-block replacement at layer 1 for $n_q=8$ and $L=4$: test and in-sample perplexity, given as the mean and sample standard deviation across three training seeds, together with top-1 agreement with the unmodified model at seed 42.
    The identity-skip row bypasses the target layer entirely without training and has no seed variance.
    The no-circuit variant retains the constrained mixers, the bounding nonlinearity and the residual gate, removes all $32$ quantum heads and the learned readout, and is trained identically to the full block.
    All values are token perplexity, for which lower is better.}
    \label{tab:single_block_ppl}
    \begin{ruledtabular}
    \begin{tabular}{lcccc}
    Condition & WikiText-2 & C4 & webmix-recon & top-1 agreement \\
    \colrule
    Baseline (unmodified)   & 8.672            & 14.305            & 8.120            & --    \\
    Identity-skip           & 11.941           & 18.706            & 11.079           & 0.800 \\
    No-circuit (mixers only)  & $12.787\pm0.838$ & $18.055\pm0.839$  & $10.640\pm0.677$ & 0.812 \\
    Full multi-head QNN                & $9.309\pm0.079$  & $15.099\pm0.104$  & $8.436\pm0.061$  & 0.856 \\
    Matched MLP             & $9.008\pm0.032$  & $14.315\pm0.032$  & $7.964\pm0.026$  & 0.853 \\
    \end{tabular}
    \end{ruledtabular}
    \end{table*}

    \begin{table*}[t]
    \setlength{\tabcolsep}{3pt}
    \caption{%
    Single-block replacement at layer 1: downstream task accuracy and validation best perplexity, each given as the mean and sample standard deviation (SD) across three training seeds.
    LAMBADA, WinoGrande and BoolQ report \texttt{acc} (accuracy), while HellaSwag, ARC-easy (ARC-e), ARC-challenge (ARC-c) and PIQA report \texttt{acc\_norm} (length-normalized accuracy).
    Validation best perplexity is the checkpoint-selection metric of Appendix~\ref{annex:uncertainty} and is shown for reference only.
    The dispersion quoted here is optimization variability across seeds and is separate from evaluation-set noise, for which the \texttt{lm-eval} bootstrap standard error ranges from $0.004$ to $0.015$ across tasks.
    Baseline and identity-skip are single runs and carry no seed variance.
    No multiple-comparison correction is applied across the seven tasks.}
    \label{tab:single_block_downstream}
    \setlength{\tabcolsep}{2.5pt}
    \begin{ruledtabular}
    \begin{tabular}{lccccccc}
    Condition & LAMBADA & HellaSwag & ARC-e & ARC-c & PIQA & WinoG & BoolQ \\
    \colrule
    Baseline        & 0.711 & 0.798 & 0.835 & 0.628 & 0.805 & 0.728 & 0.885 \\
    Identity-skip   & 0.653 & 0.781 & 0.803 & 0.599 & 0.781 & 0.663 & 0.881 \\
    No-circuit          & $0.653\pm0.004$ & $0.779\pm0.001$ & $0.817\pm0.020$ & $0.622\pm0.017$ & $0.788\pm0.004$ & $0.654\pm0.023$ & $0.877\pm0.003$ \\
    Full multi-head QNN        & $0.668\pm0.006$ & $0.784\pm0.002$ & $0.841\pm0.006$ & $0.644\pm0.011$ & $0.800\pm0.003$ & $0.715\pm0.002$ & $0.833\pm0.009$ \\
    Matched MLP     & $0.684\pm0.004$ & $0.785\pm0.002$ & $0.842\pm0.005$ & $0.633\pm0.004$ & $0.803\pm0.000$ & $0.711\pm0.005$ & $0.830\pm0.006$ \\
    \colrule
    & \multicolumn{7}{c}{Validation best PPL, mean and sample SD at $n=3$ seeds} \\
    Identity-skip   & \multicolumn{7}{c}{$8.9535$, single run without fine-tuning} \\
    No-circuit          & \multicolumn{7}{c}{$8.5152 \pm 0.3727$} \\
    Full multi-head QNN        & \multicolumn{7}{c}{$7.0081 \pm 0.0372$} \\
    Matched MLP     & \multicolumn{7}{c}{$6.6473 \pm 0.0118$} \\
    \end{tabular}
    \end{ruledtabular}
    \end{table*}

    \subsection{Disjoint multi-block replacement}

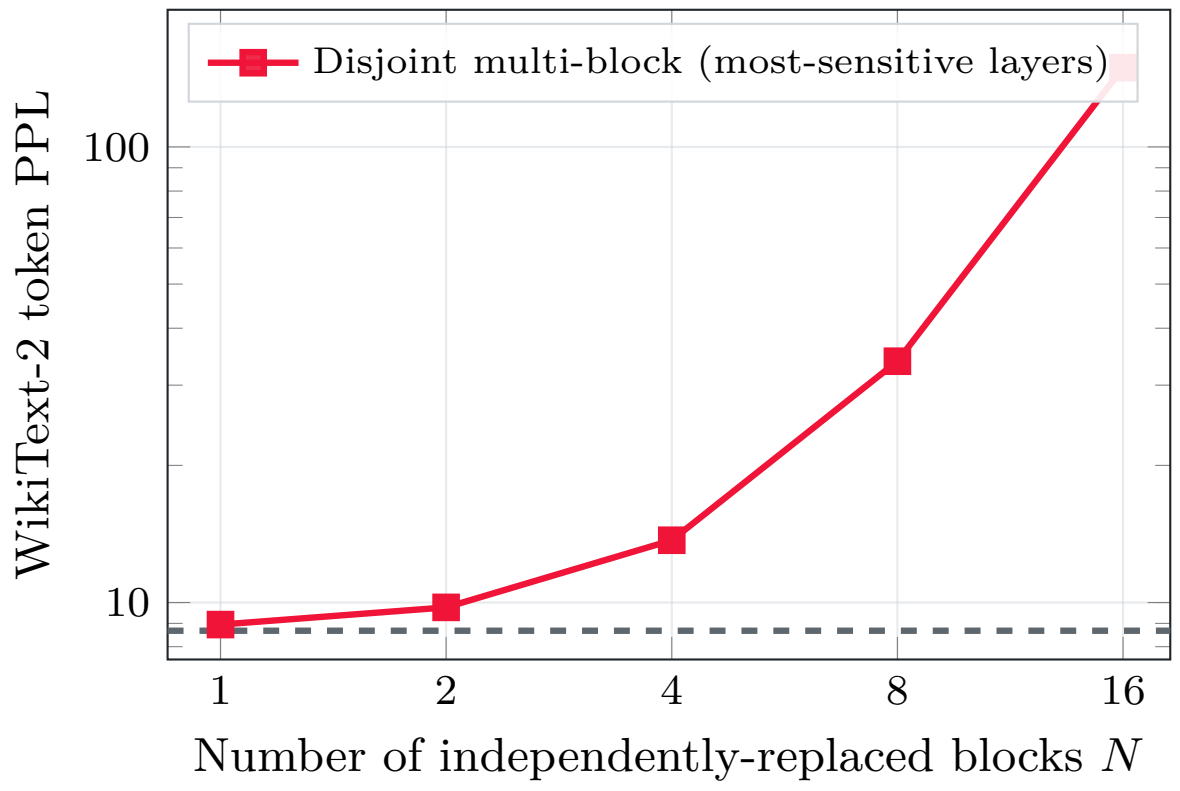
\begin{figure}[t]
        \centering
        \begin{tikzpicture}
        \begin{axis}[
            width=0.95\linewidth,
            height=0.68\linewidth,
            xmode=log,
            ymode=log,
            log ticks with fixed point,
            xlabel={Number of independently-replaced blocks $N$},
            ylabel={WikiText-2 token PPL},
            xtick={1,2,4,8,16},
            xticklabels={1,2,4,8,16},
            xmin=0.85, xmax=18.5,
            ymin=7.5, ymax=200,
            grid=major,
            grid style={mvcgray500, opacity=0.5},
            tick label style={font=\footnotesize},
            label style={font=\small},
            axis line style={mvcgray900},
            every axis plot/.append style={line width=1.2pt},
            error bars/y dir=both,
            error bars/y explicit,
            legend style={
                at={(0.02,0.98)}, anchor=north west,
                font=\scriptsize, draw=mvcgray500,
                fill=white, fill opacity=0.9, text opacity=1,
            },
            legend columns=1,
        ]

        \addplot[mvcgray700, dashed, no marks, forget plot]
            coordinates {(0.85,8.672) (18.5,8.672)};

        \addplot[mvcrose, mark=square*, mark size=2pt]
            coordinates {
            (1,8.946) +- (0,0.271)
            (2,9.758) +- (0,0.285)
            (4,13.700) +- (0,0.391)
            (8,33.861) +- (0,0.922)
            (16,149.252) +- (0,4.437)
        };
        \addlegendentry{Disjoint multi-block (most-sensitive layers)}

        \end{axis}
        \end{tikzpicture}
        \caption{%
        WikiText-2 token perplexity against the number of independently-replaced blocks $N$, at scattered depths among the model's most sensitive non-boundary layers, jointly fine-tuned.
        Trainable parameters scale as $N \times 145\,281$, unlike the fusion setting.
        Error bars show $95\%$ bootstrap confidence intervals; the dashed gray line marks the unmodified base model.
        The $N=1$ point ($8.946$) and the single-block result of Table~\ref{tab:single_block_ppl} ($9.309\pm0.079$) come from different training procedures, the former warm-started and jointly fine-tuned even at $N=1$, and should not be read as two measurements of one quantity (Appendix~\ref{annex:multiblock}).}
        \label{fig:multiblock}
    \end{figure}

    In this subsection, we test whether our block replacement composes by replacing $N$ scattered transformer blocks with $N$ independent multi-head QNN blocks and optimizing the latter jointly. The results are shown in Fig.~\ref{fig:multiblock}.
    Trainable parameters here scale as $N$ times $145\,281$, unlike the fusion setting, where we replaced with only a single multi-head QNN. Per-$N$ perplexity, downstream accuracy and layer sets are tabulated in Appendix~\ref{annex:multiblock}.

    Each substitution individually succeeds.
    At every $N$ the multi-replacement model is better than identity-skip, and the margin widens sharply with $N$, reaching $149.252$ against $1.19\times10^{5}$ on WikiText-2 at $N=16$.
    What fails is composition.
    Absolute performance degrades faster than linearly, from $8.946$ at $N=1$ to $13.700$ at $N=4$ and $149.252$ at $N=16$, while top-1 agreement falls from $0.851$ to $0.430$ and LAMBADA accuracy collapses to $0.089$.
    Replacements that are separately viable are therefore not jointly viable at this scale, at least under joint fine-tuning from independently warm-started blocks. Besides the missing token mixing, a likely, untested contributor is distribution shift: Stage 1 fits each module to activations of the unmodified model, which downstream modules no longer see.

    We present this as separate and weaker evidence rather than as a controlled contrast with block fusion.
    The scattered layer set is the head of the identity-skip sensitivity ranking with the two boundary layers excluded, running from layer 1 through the tail layers 38 to 46 and the early layers 2 to 6 and 8, whereas the fusion span is contiguous and occupies the least sensitive middle of the stack, and their identity-skip baselines differ by four orders of magnitude at $N=16$.
    Whatever separates the two families is accordingly dominated by which layers were removed rather than by the mechanism of removal, and the corresponding figures should not be directly compared against one another.

    \subsection{Execution on quantum hardware}
    \label{sec:hardware}

    Every result above was obtained with the quantum circuits simulated classically.
    To establish that the trained circuits also run on a present-day quantum device as compiled, we executed them for the five fusion checkpoints of Sec.~\ref{sec:fusion}, $N \in \{1,2,4,8,16\}$, on \texttt{ibm\_basquecountry}, a $156$-qubit IBM Heron r2 processor.
    No parameter was retrained, rescaled or recalibrated for this step: the circuits submitted are the ones whose perplexities appear in Fig.~\ref{fig:fusion_ablation}.

    Only the per-head map from a $64$-dimensional input chunk to its $64$ Pauli expectation values is executed on the device; the constrained mixers, the bounding nonlinearity, the learned readout and the residual gate remain in floating point, in the same arrangement as in deployment.
    Expectation values are estimated with \texttt{EstimatorV2} at $8\,192$ shots per circuit and resilience level $1$, that is, twirled readout error extinction (TREX) with no zero-noise extrapolation, since gate twirling and extrapolation are not available in combination with the native two-qubit rotation we rely on below.
    Before any device time was spent, the compiled circuits and the observable ordering were checked against the trained statevector simulator and agreed to $\sim\!10^{-5}$ in the expectation values, which validates the translation, not the noise model.

    The open $R_{zz}$ nearest-neighbour chain of Sec.~\ref{sec:circuit} is what makes this cheap to run on a quantum device.
    Each head carries $(n_q-1)L = 28$ two-qubit rotations and no wraparound edge, so its interaction graph is an $8$-qubit path.
    Opting the compilation target into the processor's fractional gate set makes $R_{zz}(\theta)$ native, and the $28$ logical rotations then map one-to-one onto $28$ device two-qubit gates at two-qubit depth $13$ and total instruction-set-architecture (ISA) depth $42$, with zero routing swaps.
    Without that basis each $R_{zz}$ decomposes into two \texttt{CZ}s and the two-qubit count doubles; a closed ring would in addition require the router to break the chain.

    The heads never interact, each acting on its own $64$-dimensional chunk of the activation, so several can be executed side by side in a single circuit; the $8$-qubit path shape is what leaves the device room for many of them at once.
    We pack $14$ heads onto vertex-disjoint $8$-qubit paths of the coupling map, using $112$ of the $156$ physical qubits, and the packed circuit carries $392$ two-qubit gates at an unchanged two-qubit depth of $13$ with zero routing swaps, its total ISA depth rising only from $42$ to between $45$ and $46$.
    Packing is therefore nearly free in circuit depth and, in the noiseless case, exact in outcome, since disjoint patches reproduce the sequential per-head expectation values. On hardware, crosstalk between packed patches could matter; we did not compare packed with isolated execution.
    The $k=32$ heads of one token are covered by three circuits in a $14/14/4$ split, on $112$, $112$ and $32$ qubits respectively, and one such submission occupies the processor for roughly $1.5$ minutes.

    Table~\ref{tab:hardware} reports the compiled cost at each $N$.
    It is essentially constant, which is the expected consequence of the block having a fixed size: fusing a wider span changes the trained angles but not the circuit, so a $16$-layer replacement costs the device essentially what a single-layer one does.
    This is the sense in which the fusion result matters for hardware, and it is why we regard the width of the absorbed span rather than the width of the register as the axis to push at fixed device capability.

    What the device runs for the five checkpoints is the quantum heads of the first prompt token, after which the block continues classically for the remaining decoding steps; for $N=8$ we also ran the block at every decoding step (below).
    Computing perplexity through the quantum processor is out of reach by a wide margin: at three circuits per token, the WikiText-2 test split alone would require of order $10^{6}$ submissions, against $15$ for the first-token runs and $135$ ($3$ runs $\times$ $15$ tokens $\times$ $3$) for the $N=8$ campaign.
    As a measured output, the five fusion models with the hardware expectation values substituted for the first token answer the prompt ``What is the capital of Spain?'' with ``The capital of Spain is Madrid.''; a single prompt is a weak test. The perplexities of Fig.~\ref{fig:fusion_ablation} are accordingly simulated numbers throughout, and what the hardware runs add is that the circuits behind them execute at the stated depth, gate count and qubit budget, not that they were executed for every token of the evaluation sets.

    We also run a simple, hardware-informed noise check on our trained models. We approximate the device as a global depolarizing channel, which damps every Pauli expectation the trained circuits output by a single factor derived from the median gate errors of \texttt{ibm\_basquecountry} and the per-head gate counts of the deployed ansatz (Sec.~\ref{sec:circuit}), with expectation values sampled at the same $8\,192$ shots. At $N=8$, on WikiText-2 (\texttt{lm-eval} word-level perplexity), the noisy replacement is 6.4\% worse in perplexity than the noiseless simulation and 1.4\% better than deleting the same span, so the circuits retain part of their function at this noise budget, though the margin over identity-skip narrows substantially. The model is coarse --- one factor for all heads and observables, no coherent errors or crosstalk --- so it is indicative only. Per-head deviations were measured for the $N=8$ checkpoint, whose block was run on the device at every decoding step rather than only the first: $15$ generated tokens for each of two prompts, one of them in two independent sessions ($3$ runs, $32$ heads $\times$ $64$ observables per token). Against the noiseless statevector values at the same inputs, the hardware expectation values deviate by $0.051$--$0.054$ on average, including the $8$ constant identity terms per head that cannot deviate ($95$th percentile $0.157$--$0.165$, maximum $1.03$), with per-head means between $0.027$ and $0.150$ and roughly constant along the sequence. The deviation is as large for non-identity single-qubit as for two-qubit observables ($0.059$--$0.063$ against $0.058$--$0.061$) and about six times the shot-noise scale of $0.0088$ at $8\,192$ shots, so gate and readout error, not sampling, dominate. The density-matrix simulation, which uses the gate-specific and readout errors of the device's calibration snapshot rather than a single damping factor, gives the order of magnitude but not the size (mean deviation $0.015$, a factor $3.4$--$3.6$ below the device) and only weakly the pattern across observables (correlation $0.14$--$0.15$). The depolarizing model of the check above, which multiplies every expectation by $0.901$ (a depolarizing strength of $0.099$), implies a mean deviation of $0.026$ on the same inputs, about half the measured value, which suggests that check is optimistic; the top-1 agreement below is the more direct evidence. Through the model, the top-1 prediction agrees with the unmodified model at $15$ of $15$ positions on one prompt and $13$ of $15$ on the other, as for the noiseless block, whereas knocking out the circuit outputs of the trained block (replacing them by the head's own bounded input chunk, with mixers and readout kept) keeps $1$ and $0$; the mean Kullback--Leibler divergence from the unmodified model, teacher-forced on the hardware tokens, is $0.14$ (noiseless $0.04$) on the first prompt and $0.81$ and $1.15$ (noiseless $0.71$) on the second. The two sessions on the same prompt differ by $0.34$, comparable to the gap to the noiseless block, so we draw no ranking from them. A systematic study of noise propagation beyond these runs is left to future work.

    \begin{table*}
    \caption{%
    Compiled cost of one multi-head QNN head bank on \texttt{ibm\_basquecountry} ($156$-qubit Heron r2), for the fusion checkpoint absorbing a span of $N$ layers, at $8\,192$ shots and resilience level $1$ with the native fractional $R_{zz}$ enabled.
    Each column describes one packed group of $14$ heads, the largest and most demanding of the three; the $k=32$ heads of a token are covered by three such groups in a $14/14/4$ split on $112/112/32$ qubits.
    The checkpoints are the ones whose perplexities are reported in Fig.~\ref{fig:fusion_ablation}; nothing was retrained for this step.}
    \label{tab:hardware}
    \begin{ruledtabular}
    \begin{tabular}{lrrrrr}
    & $N{=}1$ & $N{=}2$ & $N{=}4$ & $N{=}8$ & $N{=}16$ \\
    \colrule
    Total ISA depth        & 45 & 46 & 46 & 45 & 45 \\
    Two-qubit depth        & 13 & 13 & 13 & 13 & 13 \\
    Two-qubit gates        & 392 & 392 & 392 & 392 & 392 \\
    Total gates            & 3020 & 2960 & 3009 & 3002 & 3008 \\
    Physical qubits        & 112 & 112 & 112 & 112 & 112 \\
    Routing swaps          & 0 & 0 & 0 & 0 & 0 \\
    \end{tabular}
    \end{ruledtabular}
    \end{table*}

    \section{Discussion}
    \label{sec:discussion}

    \subsection{What the ablations of Secs.~\ref{sec:results}\,A--C establish}
    \label{sec:disc_ablations}

    Hybrid quantum--classical models are vulnerable to a specific criticism: that the classical components surrounding the circuit are expressive enough to solve the task alone, leaving the quantum module as decoration~\cite{cerezo2022challenges}.
    The multi-head QNN was designed against this failure, and the single-block ablation tests it directly.
    The manner of the no-circuit control's failure is as informative as the $3.5$-perplexity-point magnitude: the gate it learns collapses toward the vanishing limit ($\alpha = 0.0056$ and $0.0042$ at two of three seeds, against $0.16$--$0.18$ for the full block at every seed), so given only the mixers, optimization converges to identity-skip rather than learning a replacement for it, and its test perplexity on WikiText-2 lands above the identity-skip reference at every seed (Sec.~\ref{sec:single}, Table~\ref{tab:annex_single_seeds}).
    The mixers alone are therefore not a substitute for the heads they surround.

    Three further controls bear on what it is about the heads that matters (Appendix~\ref{annex:controls}, Table~\ref{tab:annex_controls}; seed 42, with one repeat).
    Setting every entangler angle to zero and freezing it, so that each head is a product state, gives $8.994$ on WikiText-2 and $14.369$ on C4 at seed 42, against $9.230$ and $15.001$ for the full block ($8.994$ and $9.002$ at seeds 42 and 123, against $9.230$ and $9.309$).
    Freezing every circuit parameter at its initial values, where the variational and entangling angles are small, gives $9.039$ and $14.389$, and freezing it at uniformly random angles gives $9.077$ and $14.449$; in all three cases the readout, mixers and gate are trained.
    At seed 42 all three lie below the full block by more than the paired-bootstrap interval of the evaluation documents, and within $0.02$--$0.11$ perplexity points of the parameter-matched MLP.
    Taken at face value, the single-block recovery does not require trained or entangled circuits: a fixed circuit followed by a learned linear readout recovers as much.
    These controls are single-seed apart from the entangler-off repeat, and the full block's three seeds span $9.230$--$9.387$ on WikiText-2, so we read them as removing the claim that the benefit is specific to trained entangled circuits, not as a ranking of the controls against each other or against the full block.

    A second finding concerns where the parameters sit relative to where the capability sits, and the two do not coincide.
    The learned readout accounts for $90\%$ of the block's trainable parameters at $m=d_h$ and purchases $1.5\%$ in perplexity on WikiText-2 at $N=4$, and $0.8\%$ on C4.
    This suggests a reading of its role different from the one that motivated it: its function is not to supply nonlinearity, which the encoding, the circuit and the measurement already provide, but to align the measured Pauli expectations with the basis the surrounding network expects at that point in the residual stream.

    \subsection{Comparison with the parameter-matched MLP}

    A parameter-matched MLP outperforms the full block at single-block depth by between $2.9$ and $6.7\%$ and leads it at nearly every span size in the block fusion setting.
    Therefore, the quantum module buys no accuracy over a classical module of the same parameter budget under identical training, and any future version of this construction has to close that gap.
    The comparison bounds the cost of this construction relative to a parameter-matched MLP, at a few percent; the margin by which a mature classical model leads a construction forced through an eight-qubit encoding bottleneck is the quantity to watch as circuit design, encoding strategies and hardware develop~\cite{larocca2025barren}.

    Our results establish competitiveness with a classical module of the same parameter budget, not with the best available classical compression. We hold parameter count fixed while varying only whether the computation is quantum, whereas practical compression combines pruning or layer removal with healing, quantization and tensor-network factorization~\cite{men2024shortgpt,gromov2024unreasonable,chen2025streamline,frantar2023sparsegpt,frantar2023gptq,tomut2024compactifai}, which we do not compare against.

    \subsection{Limitations}

    The multi-head QNN does not mix across tokens, and this is the limit that matters.
    It is a position-wise map, but the transformer block it replaces contains self-attention, so every replacement deletes all cross-token mixing at that depth (Sec.~\ref{sec:results}).
    The surrounding layers carry contextual information past the replaced depth, but only up to a point: the signature appears in the fusion saturation at large $N$, in the failure of scattered replacements to compose, and downstream, where layer 1 recovers only $0.015$ of the $0.058$ that identity-skip forfeits on LAMBADA, the task most dependent on long-range context, against $0.052$ of $0.065$ on WinoGrande.
    Any extension to wider or more numerous replacements will have to address token mixing directly.

    The choices that make the module trainable and executable (small registers, shallow depth, independent heads) also keep it classically simulable~\cite{cerezo2025simulability}, and all training here uses exact statevector simulation, which would not be available beyond that regime. The hardware run therefore demonstrates executability, not advantage, and its cost grows with tokens $\times$ heads $\times$ layers $\times$ shots. We did not benchmark the wall-clock cost of simulating the module against the block it replaces.

    Several controls are missing. The no-circuit control removes the circuits and the readout together, so it shows that the slot needs capacity and nonlinearity, not that this capacity must be quantum; in fusion it comes within $1$--$3\%$ of the full module in perplexity. We ran three controls on the circuits themselves (Appendix~\ref{annex:controls}): an entangler-off, product-state head, and circuits frozen at their small initial angles and at uniformly random angles; all beat the full block at seed 42, so the benefit we report cannot be attributed to trained entangled circuits. A classical trigonometric head of matched parameter count was not run, and the controls were not repeated over three seeds (the entangler-off control has two). Classical surrogates are known for several classes of parameterized-circuit models~\cite{landman2023classically,schreiber2023classical}, which motivates the trigonometric-head control. Identity-skip is untrained, so a fine-tuned base model and a healing baseline are also missing, as is a replacement of the MoE feed-forward network alone, which holds $97\%$ of a layer's parameters and would keep attention. We leave these to future work.

    Two further limits apply to the evidence.
    Only the headline configurations were trained at more than one seed, so the width, depth, fusion-ablation and multi-block comparisons cannot separate a real effect from run-to-run variation.
    And when we say a replacement matches the unmodified model, we mean that across multiple metric evaluations they are nearly the same, not exactly the same.
    Appendix~\ref{annex:caveats} lists the remaining caveats, of which the most important is the training asymmetry of Sec.~\ref{sec:results}.

    \section{Conclusions}
    \label{sec:conclusions}

    We asked whether a quantum module can be installed into a production-scale language model, trained to take over the function of the transformer blocks it displaces, evaluated against the unmodified model, and then executed on hardware that exists.
    At the scale we tested, it can.
    A multi-head QNN block of $145\,281$ trainable parameters, replacing a contiguous span of eight transformer blocks in Qwen3-30B-A3B, removes $16.33\%$ of the model's parameters and stays within about $1\%$ of the original model in perplexity on both held-out sets; the quantum heads within it (circuits and readout) are load-bearing, costing $3.5$ perplexity points when deleted at a sensitive depth (Table~\ref{tab:annex_readout} separates the two), although a product-state or frozen circuit followed by the learned readout recovers at least as much (Appendix~\ref{annex:controls}); and the trained circuits behind those numbers compile onto a $156$-qubit Heron processor at two-qubit depth $13$ with zero routing swaps, $14$ heads at a time on $112$ qubits.

    The multi-head QNN acts on each token independently, so every replacement deletes the cross-token mixing at that depth and the surrounding attention layers must carry it instead. How much of that can be removed before the model degrades is what bounds the method.


    What would have to change for this construction to be preferable rather than merely
    feasible? Executing the quantum heads is not the obstacle. Two other things are. The first is
    throughput: the $k=32$ heads of one token do not fit on the device at once and are covered in a $14/14/4$ split, so every token costs three submissions, and
    inference through the device is many orders of magnitude slower than the classical
    forward pass it replaces. The second is register width: widening $n_q$ runs into
    gradient pathologies that grow exponentially with the qubit
    count~\cite{mcclean2018barren,larocca2025barren}, and entangling the heads, the design lever that would take the module beyond the classically simulable regime, would likely raise the same concern. Circuit depth, by contrast,
    remains open for trainability (the depth limit of Sec.~\ref{sec:method} concerns hardware fidelity): better ans\"atze, native gate sets and error mitigation all have room
    to work there.

    The results point to three further directions. The first is a replacement that carries
    some analogue of attention, which would remove the sharpest structural limit we
    encountered. The second is a comparison against real compression methods rather than a
    matched control. The third is an account of why replacements that work individually
    fail to compose, which we leave open here. Whether a quantum module could also support interpretability, for example through quantities with no classical counterpart, is a question we do not address.

    What this work establishes is narrower. A quantum module can be carried end to end
    (installed, trained, evaluated, and executed) inside a production-scale language model,
    and we have marked where the boundaries of that lie. Whether such a module becomes
    preferable to a classical one of the same size depends on the circuit designs and the
    hardware that follow.

    \begin{acknowledgments}
    We acknowledge the Donostia International Physics Center (DIPC), Ikerbasque Foundation for Science, Basque Government, Diputaci\'{o}n de Gipuzkoa, European Innovation Council (EIC), Tecnun and Spanish Government for constant support. We thank the IBM Quantum Network \& BasQ for access to IBM Quantum System Two (\texttt{ibm\_basquecountry}, a Heron r2 processor). We also acknowledge Mitacs Canada (grant IT37546) for the award and support for M.J.A. This research has been supported in part through the Quantum Software Consortium (QSC), under grant \#ALLRP587590-23 from the Natural Sciences and Engineering Research Council of Canada (NSERC) Alliance Consortia Quantum Grants program.
    \end{acknowledgments}

    \bibliography{references}

    \appendix
    \renewcommand{\appendixname}{Appendix}
    \renewcommand{\thesection}{A\arabic{section}}
    \renewcommand{\thesubsection}{A\arabic{section}.\arabic{subsection}}
    \makeatletter
    \renewcommand{\p@subsection}{}
    \@removefromreset{equation}{section}
    \makeatother
    \renewcommand{\thetable}{A\Roman{table}}
    \renewcommand{\thefigure}{A\arabic{figure}}
    \renewcommand{\theequation}{A\arabic{equation}}
    \setcounter{section}{0}
    \setcounter{table}{0}
    \setcounter{figure}{0}
    \setcounter{equation}{0}

    \begin{center}
    {\Large\bfseries Appendices}
    \end{center}
    \medskip

    The appendices report the complete per-run data behind the results of the main text,
    together with the training and evaluation protocol in full, the statistical
    resolution status of every claim we make, and the caveats that attach to specific
    numbers.
    All values were recomputed from checkpoint state dictionaries, training logs,
    benchmark output, and the base-model tensor headers; configuration was read from
    the saved checkpoints rather than inferred from run names.
    Every number reported in the main text and in these appendices comes from a single
    two-stage training configuration, identical across every reported arm: Stage 1 trains at learning rate $10^{-3}$ with batch size $256$ on recorded
    activations, and Stage 2 trains at learning rate $10^{-4}$ with batch
    size $1$ for $20\,000$ steps.
    No result obtained under any other optimizer setting is reported or used anywhere
    in this work.

    \section{Model, installation, and parameter accounting}
    \label{annex:model}

    \subsection{Base model}
    All compression figures in the main text are computed on the total-parameter basis of Table~\ref{tab:annex_base}.
    Replacing a decoder layer bypasses its attention, its sparse mixture-of-experts feed-forward block, and both of its RMSNorms, removing the full $623\,120\,640$ parameters of that layer.

    \begin{table*}[!t]
    \caption{Parameter accounting for the base model, Qwen3-30B-A3B-Instruct-2507.}
    \label{tab:annex_base}
    \begin{ruledtabular}
    \begin{tabular}{lr}
    Property & Value \\
    \colrule
    Total parameters                                   & $30\,532\,122\,624$ \\
    Decoder layers                                     & $48$ \\
    Parameters per decoder layer                       & $623\,120\,640$ \\
    \quad of which MoE experts ($128$ experts)         & $603\,979\,776$ \\
    \quad of which attention ($q,k,v,o$ proj.\ and $q,k$ norm) & $18\,874\,624$ \\
    \quad of which MoE router                          & $262\,144$ \\
    \quad of which two RMSNorms                        & $4\,096$ \\
    Active parameters per layer (top-$8$ of $128$)     & $56\,889\,600$ \\
    Active parameters, whole model                     & $3\,353\,032\,704$ \\
    Non-layer parameters (embedding, LM head, final norm) & $622\,331\,904$ \\
    \end{tabular}
    \end{ruledtabular}
    \end{table*}

    \subsection{Installation and interface}
    Each replacement is installed as a hook on one decoder layer. Writing $g_{\boldsymbol{\Theta}}$ for the multi-head QNN block of Eq.~(\ref{eq:mhqnn}) of the main text, the installed map is
    \begin{equation}
    \mathbf{y} = \mathbf{x} + \alpha\,\mathcal{S}^{-1}\!\left(g_{\boldsymbol{\Theta}}\!\left(\mathcal{S}(\mathbf{x})\right)\right),
    \label{eq:annex_install}
    \end{equation}
    where $\mathcal{S}$ and $\mathcal{S}^{-1}$ are fixed min--max affine maps stored as non-trainable buffers ($4 \times 2048$ floats), recovered once during Stage 1 and held constant thereafter.
    They rescale the activation into the range the encoding contract requires and undo that rescaling on the way out; they carry no trainable parameters and are excluded from all parameter counts.
    The residual gate $\alpha$ is a single scalar parameter, and $\alpha=0$ reproduces identity-skip exactly, which is the property the vanishing-gate argument of the main text relies on.

    \subsection{Architecture conditions and parameter composition}
    Table~\ref{tab:annex_conditions} lists every condition reported in the main text, with its forward path and trainable parameter count, and Table~\ref{tab:annex_paramgroups} decomposes those counts by component.
    The variational circuit tensors proper --- the single-qubit rotation angles and the entangler angles --- account for only $1\,920$ parameters, or $1.3\%$ of the full block.
    Table~\ref{tab:annex_paramgroups} also carries the register-width and upload-depth variants of an exploratory grid. An \emph{upload pass} encodes each of a head's $d_h=64$ input features once, so a circuit with $p$ passes has $L=p\,d_h/(2n_q)$ layers and the relation $d_h=2n_qL$ holds only for a single pass, as in the deployed configuration ($n_q=8$, $p=1$). Since those cells are single-seed and differ in observable-pool composition as well as in register width, no width or depth claim is drawn from them; the deployed configuration was selected on the grounds of circuit executability set out in the main text.
    Counts exclude the Pauli-index selection buffer, of shape $(32, 64)$, which fixes which $64$ observables each head measures and is architecture rather than a fitted quantity.

    \begin{table*}[h]
    \caption{%
    Architecture conditions. Trainable counts exclude frozen buffers, including the fixed Sylvester--Hadamard factor, the min--max interface maps of Eq.~\eqref{eq:annex_install}, and the Pauli-index selection buffer.
    Mixer denotes the constrained mixer $\boldsymbol{\Lambda}_2\mathbf{H}\boldsymbol{\Lambda}_1$ of the main text.}
    \label{tab:annex_conditions}
    \begin{ruledtabular}
    \begin{tabular}{llr}
    Condition & Forward path & Trainable \\
    \colrule
    Full multi-head QNN                & mixer $\to \tanh \to$ QNN bank $\to$ learned readout $\to$ mixer & $145\,281$ \\
    Circuit-only            & as above, learned readout removed; raw expectations pass through & $14\,209$ \\
    Frozen-readout          & as full, readout held at its $0.1\!\cdot\!\mathrm{randn}$ initialization & $14\,209$ \\
    LayerNorm readout       & as full, learned readout replaced by per-head LayerNorm & $18\,305$ \\
    No-circuit (mixers only)  & mixer $\to \tanh \to$ identity $\to$ mixer; all $32$ heads removed & $8\,193$ \\
    Matched MLP             & $2048\!\to\!34 \to \tanh \to 34\!\to\!34 \to \tanh \to 34\!\to\!2048$ & $142\,537$ \\
    Identity-skip           & $\mathbf{y}=\mathbf{x}$; additive contribution zeroed at the hook & $0$ \\
    No entangler               & as full, every $R_{zz}$ angle set to $0$ and frozen (product-state heads) & $144\,385$ \\
    Frozen circuit (initial values) & as full, circuit and encoding angles frozen at their initial values & $139\,265$ \\
    Frozen circuit (uniform random) & as full, $R_x$ and $R_{zz}$ angles redrawn from $U(0,2\pi)$ and frozen & $139\,265$ \\
    \end{tabular}
    \end{ruledtabular}
    \end{table*}

    \begin{table*}[h]
    \caption{Trainable parameters by group for all reported conditions, including the exploratory register/depth grid cells and the three circuit controls (frozen groups are not trainable). ``Circuit'' is the sum of trainable encoding, single-qubit and entangler parameters. The total additionally includes the single residual scale $\alpha$, which belongs to no group.}
    \label{tab:annex_paramgroups}
    \begin{ruledtabular}
    \begin{tabular}{lrrrrrrr}
    Condition & Angles & Entangler & Encoding & Circuit & Readout & Mixers & Total \\
    \colrule
    Full multi-head QNN ($n_q=8$, $L=4$)  & $1\,024$ & $896$    & $4\,096$  & $6\,016$  & $131\,072$ & $8\,192$ & $145\,281$ \\
    Circuit-only                    & $1\,024$ & $896$    & $4\,096$  & $6\,016$  & $0$        & $8\,192$ & $14\,209$ \\
    Frozen readout                  & $1\,024$ & $896$    & $4\,096$  & $6\,016$  & frozen     & $8\,192$ & $14\,209$ \\
    LayerNorm readout               & $1\,024$ & $896$    & $4\,096$  & $6\,016$  & $4\,096$   & $8\,192$ & $18\,305$ \\
    No-circuit (mixers only)          & $0$      & $0$      & $0$       & $0$       & $0$        & $8\,192$ & $8\,193$ \\
    Matched MLP                     & --       & --       & --        & --        & --         & --       & $142\,537$ \\
    No entangler                    & $1\,024$ & frozen & $4\,096$ & $5\,120$ & $131\,072$ & $8\,192$ & $144\,385$ \\
    Frozen circuit, initial values     & frozen & frozen & frozen & $0$ & $131\,072$ & $8\,192$ & $139\,265$ \\
    Frozen circuit, uniform random     & frozen & frozen & frozen & $0$ & $131\,072$ & $8\,192$ & $139\,265$ \\
    $n_q=4$, $L=8$              & $1\,024$ & $768$    & $4\,096$  & $5\,888$  & $131\,072$ & $8\,192$ & $145\,153$ \\
    $n_q=4$, $L=16$             & $2\,048$ & $1\,536$ & $8\,192$  & $11\,776$ & $131\,072$ & $8\,192$ & $151\,041$ \\
    $n_q=4$, $L=24$             & $3\,072$ & $2\,304$ & $12\,288$ & $17\,664$ & $131\,072$ & $8\,192$ & $156\,929$ \\
    $n_q=8$, $L=8$              & $2\,048$ & $1\,792$ & $8\,192$  & $12\,032$ & $131\,072$ & $8\,192$ & $151\,297$ \\
    $n_q=8$, $L=12$             & $3\,072$ & $2\,688$ & $12\,288$ & $18\,048$ & $131\,072$ & $8\,192$ & $157\,313$ \\
    Fused block, any $N$            & $1\,024$ & $896$    & $4\,096$  & $6\,016$  & $131\,072$ & $8\,192$ & $145\,281$ \\
    \end{tabular}
    \end{ruledtabular}
    \end{table*}

    \section{Training and evaluation protocol}
    \label{annex:protocol}

    Table~\ref{tab:annex_training} states the shared training configuration of every reported run and the separation between training, validation and evaluation data, and Table~\ref{tab:annex_baseline} the reference values of the unmodified model.
    The separation was also checked by word-level $n$-gram overlap between the training subset and the held-out sets: no $13$-gram of WikiText-2 test occurs in the training subset ($0$ of $205\,974$) and $0.005\%$ of those of C4 validation do ($17$ of $361\,957$, in $3$ of $999$ documents); the overlap of $8$-grams is $0.02\%$ for both sets. The full Wikipedia pool does contain $9.0\%$ of the WikiText-2 $13$-grams.

    \begin{table*}[h]
    \caption{Training protocol and data separation. All primary and ablation runs share this configuration. Only \texttt{webmix-recon} overlaps the training corpus; it is reported as an in-corpus descriptive quantity throughout and never as evidence of generalization.}
    \label{tab:annex_training}
    \begin{ruledtabular}
    \begin{tabular}{ll}
    Parameter & Value \\
    \colrule
    Stage 1 & Local reconstruction, per-layer input/output fit \\
    Stage 2 & Full-model causal-LM optimization of the replacement block \\
    Objective & Stage 1: $\mathrm{MSE}-0.1\cos$ reconstruction loss; Stage 2: next-token negative log-likelihood \\
    Training corpus & \texttt{webmix\_calib}, up to $10\,000$ sequences \\
    Validation & $256$ sequences from the disjoint tail, no gradients \\
    In-corpus evaluation & \texttt{webmix-recon}, $1\,024$ sequences; overlaps the training portion \\
    Held-out evaluation & WikiText-2-raw test ($583$ sequences) and C4 English validation ($1\,024$); never loaded \\
    Stage 1 optimizer & learning rate $10^{-3}$, batch size $256$, $\leq 30$ epochs on recorded activations \\
    Stage 2 optimizer & learning rate $10^{-4}$, batch size $1$ \\
    Sequence length & $1024$ tokens \\
    Training steps (stage 2) & $20\,000$ \\
    Evaluation frequency & every $500$ steps, giving $40$ curve points \\
    Checkpoint selection & best validation perplexity over the $40$ points \\
    Training seeds & $42$, $123$, $7$ (single-block); $42$, with $123$ replication at $N=4,8$ (fusion) \\
    Observable selection seed & fixed; architecture, not a replicate \\
    Simulation & classical statevector, complex64; no device noise or sampling error \\
    Hardware & NVIDIA H200, two jobs per device with gradient checkpointing \\
    \end{tabular}
    \end{ruledtabular}
    \end{table*}

    \begin{table}[h]
    \caption{Unmodified Qwen3-30B-A3B reference values, with $95\%$ intervals for perplexity and \texttt{lm-eval} standard errors for accuracy.}
    \label{tab:annex_baseline}
    \begin{ruledtabular}
    \begin{tabular}{lr}
    Metric & Value \\
    \colrule
    WikiText-2 token PPL & $8.672\ [8.417,\,8.935]$ \\
    C4 token PPL         & $14.305\ [13.870,\,14.754]$ \\
    \texttt{webmix-recon} PPL & $8.120\ [7.834,\,8.415]$ \\
    LAMBADA              & $0.711 \pm 0.006$ \\
    HellaSwag            & $0.798 \pm 0.004$ \\
    ARC-easy             & $0.835 \pm 0.008$ \\
    ARC-challenge        & $0.628 \pm 0.014$ \\
    PIQA                 & $0.805 \pm 0.009$ \\
    WinoGrande           & $0.728 \pm 0.012$ \\
    BoolQ                & $0.885 \pm 0.006$ \\
    \end{tabular}
    \end{ruledtabular}
    \end{table}

    \subsection{Metrics and uncertainty}
    \label{annex:uncertainty}
    Two sources of uncertainty are kept separate throughout and never combined into a single figure.
    \emph{Evaluation uncertainty at fixed weights.} Perplexity intervals are document bootstraps over $1000$ resamples at a fixed bootstrap seed, computed on the negative log-likelihood (NLL) scale and exponentiated, so the intervals are asymmetric; downstream accuracy carries the \texttt{lm-eval} bootstrap standard error; top-1 agreement is a sequence bootstrap over $32$ sequences only, and its intervals are correspondingly wide.
    \emph{Optimization variability.} Reported as the sample standard deviation over three training seeds; at $n=3$ this estimates seed sensitivity, not a population confidence interval.
    \emph{Paired bootstrap.} Twelve conditions (the nine of the main campaign and the three seed-42 circuit controls) retain per-document negative log-likelihood vectors, allowing a paired bootstrap in which the same document indices are resampled for both arms, so shared document difficulty cancels; where the vectors were not retained we fall back on the Cauchy--Schwarz worst-case bound (the sum of the two standard errors), which is conservative and resolves fewer comparisons.
    Validation perplexity is the minimum over $40$ measurements on the $256$-sequence tail: a checkpoint-selection metric carrying min-of-$40$ optimism, not an unbiased test estimate, and never used for a cross-condition claim in the main text.
    Downstream tasks are LAMBADA OpenAI ($5\,153$ items, \texttt{acc}), HellaSwag ($10\,042$, \texttt{acc\_norm}), ARC-easy ($2\,376$, \texttt{acc\_norm}), ARC-challenge ($1\,172$, \texttt{acc\_norm}), PIQA ($1\,838$, \texttt{acc\_norm}), WinoGrande ($1\,267$, \texttt{acc}), and BoolQ ($3\,270$, \texttt{acc}); no multiple-comparison correction is applied across the seven tasks.

    \section{Single-block replacement: per-seed data and readout ablations}
    \label{annex:single}

    Layer 1, $n_q=8$, $L=4$, three training seeds per condition.
    Table~\ref{tab:annex_single_seeds} gives every individual run behind the seed-aggregated values of Table~\ref{tab:single_block_ppl} of the main text, and Table~\ref{tab:annex_single_agg} the aggregates themselves.
    On WikiText-2 the across-seed dispersion of the full block on the log scale is $0.00845$ against a mean within-run bootstrap standard error of $0.01526$, a ratio of $0.55$: optimization variability is roughly half the measurement noise, which is why the two are reported separately.

    The gate column of Table~\ref{tab:annex_single_seeds} supports the main-text argument about the manner of the no-circuit control's failure: two of its three seeds converge to gates of a few thousandths and land near the identity-skip reference, while the full block holds its gate between $0.157$ and $0.178$ at every seed.
    Gate magnitudes are not comparable across conditions in the nonzero regime, since the trainable diagonals of the mixers can rescale internally; only the vanishing limit carries an unambiguous reading.

    \begin{table*}[h]
    \caption{%
    Single-block replacement at layer 1, per seed.
    Validation PPL is the checkpoint-selection metric of Sec.~\ref{annex:uncertainty}.
    The gate column is the converged residual scale $\alpha$.
    Identity-skip is untrained and has no seed variance; its Val PPL entry is the validation-set value of the model with the layer bypassed.}
    \label{tab:annex_single_seeds}
    \begin{ruledtabular}
    \begin{tabular}{llrrrrrr}
    Condition & Seed & Val PPL & Gate $\alpha$ & WikiText-2 & C4 & \texttt{webmix-recon} & top-1 \\
    \colrule
    Baseline        & --  & --      & --       & $8.672$  & $14.305$ & $8.120$  & --      \\
    Identity-skip   & --  & $8.9535$& $0$      & $11.941$ & $18.706$ & $11.079$ & $0.800$ \\
    \colrule
    Full multi-head QNN        & 42  & $6.9706$& $0.1744$ & $9.230$  & $15.001$ & $8.376$  & $0.856$ \\
    Full multi-head QNN        & 123 & $7.0087$& $0.1777$ & $9.309$  & $15.087$ & $8.433$  & $0.854$ \\
    Full multi-head QNN        & 7   & $7.0451$& $0.1571$ & $9.387$  & $15.208$ & $8.497$  & $0.853$ \\
    \colrule
    Matched MLP     & 42  & $6.6605$& $0.5632$ & $8.972$  & $14.351$ & $7.989$  & $0.853$ \\
    Matched MLP     & 123 & $6.6377$& $0.5598$ & $9.021$  & $14.289$ & $7.938$  & $0.852$ \\
    Matched MLP     & 7   & $6.6437$& $0.5682$ & $9.031$  & $14.303$ & $7.965$  & $0.854$ \\
    \colrule
    No-circuit          & 42  & $8.0862$& $0.0332$ & $12.050$ & $17.091$ & $9.858$  & $0.812$ \\
    No-circuit          & 123 & $8.7004$& $0.0056$ & $13.698$ & $18.618$ & $11.022$ & $0.795$ \\
    No-circuit          & 7   & $8.7591$& $0.0042$ & $12.613$ & $18.455$ & $11.039$ & $0.798$ \\
    \end{tabular}
    \end{ruledtabular}
    \end{table*}

    \begin{table*}[h]
    \caption{Single-block aggregates corresponding to Table~\ref{tab:single_block_ppl} of the main text, mean and sample standard deviation at $n=3$ seeds.}
    \label{tab:annex_single_agg}
    \begin{ruledtabular}
    \begin{tabular}{lrrrr}
    Condition & Val PPL & WikiText-2 & C4 & \texttt{webmix-recon} \\
    \colrule
    Full multi-head QNN     & $7.0081 \pm 0.0372$ & $9.3087 \pm 0.0787$  & $15.0987 \pm 0.1038$ & $8.4356 \pm 0.0606$ \\
    Matched MLP  & $6.6473 \pm 0.0118$ & $9.0075 \pm 0.0316$  & $14.3146 \pm 0.0324$ & $7.9640 \pm 0.0255$ \\
    No-circuit       & $8.5152 \pm 0.3727$ & $12.7870 \pm 0.8375$ & $18.0548 \pm 0.8387$ & $10.6395 \pm 0.6770$ \\
    \end{tabular}
    \end{ruledtabular}
    \end{table*}

    \subsection{Readout ablations}
    Table~\ref{tab:annex_readout} isolates the learned readout at layer 1 under three separate treatments, all at seed 42 and sharing the interface, gate, schedule and seed of the full block.
    Removing the readout entirely and freezing it at a random initialization cost almost the same amount, $0.278$ and $0.257$ perplexity on WikiText-2, which is consistent with the readout performing a fitted basis alignment rather than supplying representational capacity that the rest of the block lacks; replacing it with a per-head LayerNorm, at $18\,305$ parameters, recovers most of the gap.
    This family is single-seed; removing the readout and replacing it with a LayerNorm are resolved against the full block by a paired bootstrap, while the frozen-readout arm rests on the Cauchy--Schwarz bound only.

    \begin{table*}[h]
    \setlength{\tabcolsep}{3pt}
    \caption{%
    Readout ablations at layer 1, seed 42.
    The three treatments remove the readout, freeze it at its random initialization, and replace it with a per-head LayerNorm.
    The no-circuit control is a different architecture, listed for reference.}
    \label{tab:annex_readout}
    \begin{ruledtabular}
    \footnotesize
    \begin{tabular}{lrrrrrr}
    Condition & Trainable & Val PPL & Gate $\alpha$ & WikiText-2 & C4 & \texttt{webmix-recon} \\
    \colrule
    Full multi-head QNN (reference) & $145\,281$ & $6.9706$ & $0.1744$ & $9.230\ [8.957,\,9.511]$ & $15.001\ [14.545,\,15.471]$ & $8.376\ [8.078,\,8.686]$ \\
    LayerNorm readout    & $18\,305$  & $7.0396$ & $0.3925$ & $9.332\ [9.055,\,9.618]$ & $15.094\ [14.636,\,15.568]$ & $8.466\ [8.166,\,8.777]$ \\
    Frozen readout       & $14\,209$  & $7.1555$ & $0.1844$ & $9.487\ [9.208,\,9.775]$ & $15.249\ [14.785,\,15.728]$ & $8.597\ [8.295,\,8.911]$ \\
    Circuit-only         & $14\,209$  & $7.1753$ & $0.2091$ & $9.508\ [9.228,\,9.795]$ & $15.284\ [14.819,\,15.764]$ & $8.645\ [8.342,\,8.958]$ \\
    No-circuit (mixers only)    & $8\,193$   & $8.0862$ & $0.0332$ & $12.050\ [11.325,\,12.822]$ & $17.091\ [16.504,\,17.699]$ & $9.858\ [9.509,\,10.219]$ \\
    \end{tabular}
    \end{ruledtabular}
    \end{table*}

    \subsection{Circuit controls}
    \label{annex:controls}

    Three controls test whether the single-block benefit depends on the circuits being trained and entangled.
    All use the recipe of the full block (layer 1, $n_q=8$, $L=4$, open $R_{zz}$ chain, 200\,000 local samples for 30 epochs, then $20\,000$ global steps at learning rate $10^{-4}$), and keep the mixers, the learned readout and the gate trainable.
    \emph{No entangler}: every $R_{zz}$ angle is set to zero and frozen ($896$ parameters), so each head is a product state; $144\,385$ parameters are trainable.
    \emph{Frozen circuit, initial values}: the variational $R_x$ angles, the $R_{zz}$ angles and the affine encoding angles are frozen at their initial values ($6\,016$ parameters); the variational and entangling angles start as $0.1\,\mathcal{N}(0,1)$ and the encoding weights at $\pi$, so this is a fixed encoding followed by weak rotations, not a random circuit; $139\,265$ parameters are trainable.
    \emph{Frozen circuit, uniformly random}: as the previous arm, but the $R_x$ and $R_{zz}$ angles are first redrawn from $U(0,2\pi)$ with a fixed generator (seed 42); the encoding affines keep their initial values.
    Held-out evaluation uses the same harness as Table~\ref{tab:annex_single_seeds}; the results are in Table~\ref{tab:annex_controls}.

    \begin{table*}[h]
    \caption{%
    Circuit controls at layer 1. Reference rows are those of Table~\ref{tab:annex_single_seeds}. Validation PPL for the controls is the best value of the training log on the same $256$-sequence webmix tail. All control rows are one run; the no-entangler arm was run at two seeds.}
    \label{tab:annex_controls}
    \begin{ruledtabular}
    \begin{tabular}{lrrrrr}
    Condition & Seed & Val PPL & WikiText-2 & C4 & top-1 \\
    \colrule
    Full multi-head QNN        & 42  & $6.9706$ & $9.230$ & $15.001$ & $0.856$ \\
    Full multi-head QNN        & 123 & $7.0087$ & $9.309$ & $15.087$ & $0.854$ \\
    Matched MLP                & 42  & $6.6605$ & $8.972$ & $14.351$ & $0.853$ \\
    \colrule
    No entangler               & 42  & $6.6733$ & $8.994$ & $14.369$ & $0.859$ \\
    No entangler               & 123 & $6.6659$ & $9.002$ & $14.402$ & $0.853$ \\
    Frozen circuit (initial values) & 42 & $6.6618$ & $9.039$ & $14.389$ & $0.853$ \\
    Frozen circuit (uniform random) & 42 & $6.6946$ & $9.077$ & $14.449$ & $0.856$ \\
    \end{tabular}
    \end{ruledtabular}
    \end{table*}

    Paired bootstrap over the evaluation documents ($4\,000$ resamples; seed-42 control minus seed-42 reference, in perplexity points, with $95\%$ interval; WikiText-2 then C4).
    Against the full block: no entangler $-0.236\ [-0.269,\,-0.204]$ and $-0.632\ [-0.674,\,-0.592]$; frozen circuit, initial values, $-0.191\ [-0.224,\,-0.160]$ and $-0.612\ [-0.655,\,-0.574]$; frozen circuit, uniform random, $-0.153\ [-0.182,\,-0.123]$ and $-0.552\ [-0.591,\,-0.514]$.
    Against the matched MLP: no entangler $+0.022\ [+0.004,\,+0.040]$ and $+0.018\ [-0.001,\,+0.037]$; initial values $+0.067\ [+0.049,\,+0.086]$ and $+0.038\ [+0.019,\,+0.057]$; uniform random $+0.105\ [+0.087,\,+0.123]$ and $+0.098\ [+0.079,\,+0.117]$.
    These intervals capture evaluation-set sampling only, not seed variation, which for the full block spans $9.230$--$9.387$ on WikiText-2 (Table~\ref{tab:annex_single_seeds}); per-document likelihoods were not retained for the seed-123 runs, so no paired interval is given for them.
    During training the no-entangler arm fluctuated by about $0.3$ validation perplexity points between evaluations in its last $2\,000$ steps at seed 42, so its selected checkpoint carries selection optimism of the kind described in Sec.~\ref{annex:caveats}.
    A classical trigonometric head of matched parameter count was not run.

    \section{Block fusion: complete data}
    \label{annex:fusion}

    One block of $145\,281$ trainable parameters, fixed across $N$, replaces the contiguous span from layer 10 to layer $9+N$, with all $N$ original layers bypassed.
    Table~\ref{tab:annex_fusion_ppl} is the complete numerical basis of Fig.~\ref{fig:fusion_ablation} of the main text, and additionally reports the frozen-readout arm omitted from that figure for legibility.
    Three orderings from this table are quoted in the main text.
    The full block is below the no-circuit control on all three evaluation sets at all five widths, $15$ of $15$.
    The circuit-only arm leads the no-circuit control on WikiText-2 at $N=1$, $2$ and $4$ and inverts by $0.013$ and $0.002$ at $N=8$ and $N=16$; on C4 it leads at every width.
    The matched MLP leads the full block on WikiText-2 and \texttt{webmix-recon} at every width, the two separated by only $0.002$ on C4 at $N=16$.

    \begin{table*}[h]
    \footnotesize
    \setlength{\tabcolsep}{3pt}
    \caption{%
    Fusion ablation, held-out and in-corpus perplexity with $95\%$ bootstrap intervals.
    The identity-skip rows drop the same span without supplying a replacement.
    The frozen-readout arm is reported here only; it is omitted from Fig.~\ref{fig:fusion_ablation} of the main text.}
    \label{tab:annex_fusion_ppl}
    \begin{ruledtabular}
    \begin{tabular}{llrrr}
    $N$ & Condition & WikiText-2 & C4 & \texttt{webmix-recon} \\
    \colrule
    -- & Baseline & $8.672\ [8.417,\,8.935]$ & $14.305\ [13.870,\,14.754]$ & $8.120\ [7.834,\,8.415]$ \\
    \colrule
    $1$ & Identity-skip  & $8.791$ & $14.432$ & $8.240$ \\
    $1$ & Full multi-head QNN & $8.144\ [7.903,\,8.392]$ & $13.628\ [13.221,\,14.046]$ & $7.595\ [7.321,\,7.879]$ \\
    $1$ & Frozen readout & $8.204\ [7.963,\,8.454]$ & $13.732\ [13.323,\,14.154]$ & $7.659\ [7.383,\,7.945]$ \\
    $1$ & Circuit-only   & $8.194\ [7.952,\,8.443]$ & $13.664\ [13.255,\,14.085]$ & $7.648\ [7.373,\,7.932]$ \\
    $1$ & No-circuit     & $8.315\ [8.070,\,8.567]$ & $13.853\ [13.440,\,14.278]$ & $7.800\ [7.523,\,8.088]$ \\
    $1$ & Matched MLP    & $7.969\ [7.733,\,8.212]$ & $13.323\ [12.925,\,13.733]$ & $7.371\ [7.102,\,7.651]$ \\
    \colrule
    $2$ & Identity-skip  & $8.899$ & $14.604$ & $8.362$ \\
    $2$ & Full multi-head QNN & $8.250\ [8.008,\,8.499]$ & $13.711\ [13.303,\,14.131]$ & $7.674\ [7.399,\,7.958]$ \\
    $2$ & Frozen readout & $8.257\ [8.014,\,8.508]$ & $13.765\ [13.355,\,14.186]$ & $7.717\ [7.441,\,8.004]$ \\
    $2$ & Circuit-only   & $8.277\ [8.033,\,8.529]$ & $13.820\ [13.409,\,14.243]$ & $7.782\ [7.506,\,8.069]$ \\
    $2$ & No-circuit     & $8.377\ [8.130,\,8.630]$ & $13.975\ [13.561,\,14.402]$ & $7.902\ [7.623,\,8.191]$ \\
    $2$ & Matched MLP    & $8.030\ [7.793,\,8.273]$ & $13.336\ [12.940,\,13.745]$ & $7.417\ [7.147,\,7.697]$ \\
    \colrule
    $4$ & Identity-skip  & $8.951$ & $14.848$ & $8.582$ \\
    $4$ & Full multi-head QNN & $8.314\ [8.069,\,8.567]$ & $13.913\ [13.501,\,14.338]$ & $7.836\ [7.558,\,8.123]$ \\
    $4$ & Frozen readout & $8.401\ [8.155,\,8.655]$ & $14.004\ [13.590,\,14.431]$ & $7.896\ [7.619,\,8.184]$ \\
    $4$ & Circuit-only   & $8.441\ [8.191,\,8.698]$ & $14.024\ [13.609,\,14.452]$ & $7.920\ [7.641,\,8.209]$ \\
    $4$ & No-circuit     & $8.488\ [8.239,\,8.744]$ & $14.152\ [13.736,\,14.581]$ & $8.044\ [7.763,\,8.334]$ \\
    $4$ & Matched MLP    & $8.186\ [7.945,\,8.434]$ & $13.577\ [13.175,\,13.991]$ & $7.582\ [7.309,\,7.865]$ \\
    \colrule
    $8$ & Identity-skip  & $9.659$ & $15.459$ & $9.066$ \\
    $8$ & Full multi-head QNN & $8.748\ [8.493,\,9.011]$ & $14.310\ [13.888,\,14.744]$ & $8.170\ [7.886,\,8.464]$ \\
    $8$ & Frozen readout & $8.825\ [8.569,\,9.088]$ & $14.479\ [14.053,\,14.919]$ & $8.257\ [7.971,\,8.552]$ \\
    $8$ & Circuit-only   & $8.850\ [8.592,\,9.115]$ & $14.528\ [14.100,\,14.969]$ & $8.339\ [8.053,\,8.636]$ \\
    $8$ & No-circuit     & $8.837\ [8.580,\,9.101]$ & $14.543\ [14.116,\,14.984]$ & $8.344\ [8.057,\,8.641]$ \\
    $8$ & Matched MLP    & $8.562\ [8.311,\,8.820]$ & $14.238\ [13.818,\,14.672]$ & $8.036\ [7.754,\,8.328]$ \\
    \colrule
    $16$ & Identity-skip & $13.115$ & $19.898$ & $12.129$ \\
    $16$ & Full multi-head QNN & $9.918\ [9.633,\,10.212]$ & $15.785\ [15.325,\,16.258]$ & $9.220\ [8.909,\,9.542]$ \\
    $16$ & Frozen readout & $9.902\ [9.617,\,10.194]$ & $15.843\ [15.382,\,16.317]$ & $9.245\ [8.933,\,9.567]$ \\
    $16$ & Circuit-only   & $10.091\ [9.801,\,10.391]$ & $16.084\ [15.615,\,16.566]$ & $9.400\ [9.085,\,9.727]$ \\
    $16$ & No-circuit     & $10.089\ [9.799,\,10.387]$ & $16.185\ [15.714,\,16.671]$ & $9.493\ [9.176,\,9.820]$ \\
    $16$ & Matched MLP    & $9.868\ [9.584,\,10.160]$ & $15.787\ [15.325,\,16.263]$ & $9.114\ [8.804,\,9.435]$ \\
    \end{tabular}
    \end{ruledtabular}
    \end{table*}

    \begin{table*}[h]
    \setlength{\tabcolsep}{2pt}
    \caption{%
    Fusion downstream accuracy, seed 42.
    The \texttt{lm-eval} bootstrap standard error is given in the final row and is essentially constant within a column across conditions and widths.}
    \label{tab:annex_fusion_downstream}
    \begin{ruledtabular}
    \footnotesize
    \begin{tabular}{llrrrrrrr}
    $N$ & Condition & LAMBADA & HellaSwag & ARC-e & ARC-c & PIQA & WinoG & BoolQ \\
    \colrule
    $1$ & Full multi-head QNN       & $0.706$ & $0.795$ & $0.838$ & $0.634$ & $0.813$ & $0.725$ & $0.841$ \\
    $1$ & Frozen readout & $0.700$ & $0.796$ & $0.845$ & $0.625$ & $0.812$ & $0.732$ & $0.839$ \\
    $1$ & Circuit-only   & $0.698$ & $0.799$ & $0.838$ & $0.619$ & $0.815$ & $0.739$ & $0.847$ \\
    $1$ & No-circuit         & $0.699$ & $0.798$ & $0.840$ & $0.618$ & $0.806$ & $0.725$ & $0.850$ \\
    $1$ & Matched MLP    & $0.710$ & $0.797$ & $0.827$ & $0.607$ & $0.817$ & $0.736$ & $0.824$ \\
    \colrule
    $2$ & Full multi-head QNN       & $0.711$ & $0.795$ & $0.842$ & $0.623$ & $0.815$ & $0.724$ & $0.831$ \\
    $2$ & Frozen readout & $0.699$ & $0.795$ & $0.851$ & $0.617$ & $0.820$ & $0.728$ & $0.855$ \\
    $2$ & Circuit-only   & $0.695$ & $0.796$ & $0.819$ & $0.589$ & $0.811$ & $0.721$ & $0.806$ \\
    $2$ & No-circuit         & $0.703$ & $0.792$ & $0.838$ & $0.605$ & $0.810$ & $0.727$ & $0.838$ \\
    $2$ & Matched MLP    & $0.721$ & $0.799$ & $0.829$ & $0.607$ & $0.817$ & $0.730$ & $0.840$ \\
    \colrule
    $4$ & Full multi-head QNN       & $0.688$ & $0.787$ & $0.822$ & $0.599$ & $0.814$ & $0.711$ & $0.827$ \\
    $4$ & Frozen readout & $0.688$ & $0.788$ & $0.840$ & $0.608$ & $0.818$ & $0.724$ & $0.849$ \\
    $4$ & Circuit-only   & $0.689$ & $0.788$ & $0.843$ & $0.622$ & $0.804$ & $0.717$ & $0.842$ \\
    $4$ & No-circuit         & $0.688$ & $0.785$ & $0.838$ & $0.618$ & $0.813$ & $0.712$ & $0.844$ \\
    $4$ & Matched MLP    & $0.709$ & $0.790$ & $0.823$ & $0.593$ & $0.814$ & $0.719$ & $0.798$ \\
    \colrule
    $8$ & Full multi-head QNN       & $0.683$ & $0.768$ & $0.835$ & $0.605$ & $0.802$ & $0.704$ & $0.865$ \\
    $8$ & Frozen readout & $0.664$ & $0.764$ & $0.835$ & $0.599$ & $0.801$ & $0.707$ & $0.862$ \\
    $8$ & Circuit-only   & $0.685$ & $0.766$ & $0.827$ & $0.596$ & $0.799$ & $0.707$ & $0.857$ \\
    $8$ & No-circuit         & $0.666$ & $0.768$ & $0.812$ & $0.571$ & $0.805$ & $0.700$ & $0.847$ \\
    $8$ & Matched MLP    & $0.687$ & $0.774$ & $0.823$ & $0.583$ & $0.807$ & $0.702$ & $0.845$ \\
    \colrule
    $16$ & Full multi-head QNN       & $0.637$ & $0.721$ & $0.767$ & $0.496$ & $0.783$ & $0.605$ & $0.762$ \\
    $16$ & Frozen readout & $0.627$ & $0.718$ & $0.774$ & $0.500$ & $0.787$ & $0.611$ & $0.767$ \\
    $16$ & Circuit-only   & $0.619$ & $0.719$ & $0.757$ & $0.495$ & $0.782$ & $0.617$ & $0.741$ \\
    $16$ & No-circuit         & $0.618$ & $0.718$ & $0.775$ & $0.512$ & $0.784$ & $0.612$ & $0.759$ \\
    $16$ & Matched MLP    & $0.636$ & $0.720$ & $0.790$ & $0.521$ & $0.789$ & $0.613$ & $0.784$ \\
    \colrule
    \multicolumn{2}{l}{\texttt{lm-eval} s.e.} & $0.006$--$0.007$ & $0.004$--$0.005$ & $0.007$--$0.009$ & $0.014$--$0.015$ & $0.009$--$0.010$ & $0.012$--$0.014$ & $0.006$--$0.008$ \\
    \end{tabular}
    \end{ruledtabular}
    \end{table*}

    \begin{table*}[h]
    \caption{Top-1 next-token agreement with the unmodified model on C4, with $95\%$ sequence-bootstrap intervals over $32$ sequences.}
    \label{tab:annex_fusion_fid}
    \begin{ruledtabular}
    \begin{tabular}{lrrrrr}
    $N$ & Full multi-head QNN & Frozen readout & Circuit-only & No-circuit & Matched MLP \\
    \colrule
    $1$  & $0.883\ [0.875,\,0.892]$ & $0.882\ [0.873,\,0.892]$ & $0.890\ [0.881,\,0.899]$ & $0.888\ [0.880,\,0.896]$ & $0.873\ [0.863,\,0.882]$ \\
    $2$  & $0.875\ [0.866,\,0.884]$ & $0.878\ [0.869,\,0.888]$ & $0.876\ [0.867,\,0.886]$ & $0.880\ [0.871,\,0.889]$ & $0.859\ [0.849,\,0.868]$ \\
    $4$  & $0.861\ [0.850,\,0.872]$ & $0.867\ [0.857,\,0.877]$ & $0.866\ [0.857,\,0.875]$ & $0.863\ [0.854,\,0.873]$ & $0.854\ [0.844,\,0.864]$ \\
    $8$  & $0.838\ [0.829,\,0.848]$ & $0.847\ [0.838,\,0.857]$ & $0.840\ [0.830,\,0.850]$ & $0.839\ [0.827,\,0.851]$ & $0.835\ [0.825,\,0.847]$ \\
    $16$ & $0.775\ [0.764,\,0.788]$ & $0.780\ [0.768,\,0.792]$ & $0.779\ [0.768,\,0.792]$ & $0.773\ [0.761,\,0.785]$ & $0.778\ [0.766,\,0.789]$ \\
    \end{tabular}
    \end{ruledtabular}
    \end{table*}

    Downstream accuracy (Table~\ref{tab:annex_fusion_downstream}) does not reproduce the perplexity ordering: within a given $N$ the spread across the five conditions sits at or below the standard error on HellaSwag, PIQA and WinoGrande, and where it exceeds it (BoolQ, ARC-challenge) the condition ranking is unstable across $N$, so the main text draws its component conclusions from perplexity alone.
    Degradation \emph{with} $N$, by contrast, is orderly and common to all five conditions, steepest on the tasks most dependent on long-range context.
    Top-1 agreement (Table~\ref{tab:annex_fusion_fid}) falls monotonically with $N$ for every condition, does not separate the conditions at any width, and does not track perplexity --- the circuit-only arm has the highest agreement at $N=1$ despite being worse than the full block --- which is why agreement and quality are reported as separate quantities.

    Fusion at $N=4$ and $N=8$ was replicated at a second seed (Table~\ref{tab:annex_fusion_seed}): seed 123 reproduces seed 42 to within $|\Delta\mathrm{NLL}| \leq 0.0016$ on every evaluation set (the table gives WikiText-2; on C4 and \texttt{webmix-recon} the values are at most $0.0010$ in magnitude), an order of magnitude below the bootstrap standard error, so the shape of the fusion curve is not a seed artifact (a two-point check, not a variance estimate).

    \begin{table}[h]
    \setlength{\tabcolsep}{2pt}
    \caption{%
    Fusion seed replication at $N=4$ and $N=8$; $\Delta$NLL is the WikiText-2 change relative to seed 42.
    The seed-123 logs carry no evaluation curve, so no validation perplexity is available.}
    \label{tab:annex_fusion_seed}
    \begin{ruledtabular}
    \footnotesize
    \setlength{\tabcolsep}{2pt}
    \begin{tabular}{llrrrrr}
    & & & & \texttt{webmix-} & & $\Delta$NLL \\
    $N$ & Seed & WikiText-2 & C4 & \texttt{recon} & top-1 & vs.\ s42 \\
    \colrule
    $4$ & 42  & $8.314$ & $13.913$ & $7.836$ & $0.861$ & -- \\
    $4$ & 123 & $8.307$ & $13.927$ & $7.832$ & $0.862$ & $-0.0008$ \\
    $8$ & 42  & $8.748$ & $14.310$ & $8.170$ & $0.838$ & -- \\
    $8$ & 123 & $8.762$ & $14.316$ & $8.175$ & $0.839$ & $+0.0016$ \\
    \end{tabular}
    \end{ruledtabular}
    \end{table}

    \section{Disjoint multi-block replacement}
    \label{annex:multiblock}

    Trainable parameters scale as $N \times 145\,281$, unlike the fusion setting.
    Table~\ref{tab:annex_multiblock} reports perplexity against the identity-skip reference of each layer set, and Table~\ref{tab:annex_multiblock_downstream} downstream accuracy and top-1 agreement.
    As stated in the main text, this family is not a controlled contrast with block fusion: the scattered layer set is drawn from the model's most sensitive layers, while the fusion span occupies the least sensitive middle of the stack, and the two identity-skip references differ by four orders of magnitude at $N=16$.
    The $N=1$ cell is the same architecture at the same layer as the single-block headline of Sec.~\ref{annex:single} but reaches a different value in validation perplexity ($6.4992$ against $7.0081$) because the warm start and gate initialization differ; the two are not comparable.

    \begin{table*}[h]
    \caption{%
    Disjoint multi-block replacement.
    The identity-skip column is the reference for the corresponding layer set, that is, the model with the same layers removed and nothing trained.}
    \label{tab:annex_multiblock}
    \begin{ruledtabular}
    \begin{tabular}{llrrrrrr}
    $N$ & Layer set & Trainable & Val PPL & WikiText-2 & Identity-skip & C4 & \texttt{webmix-recon} \\
    \colrule
    $1$  & $\{1\}$ & $145\,281$ & $6.4992$ & $8.946$ & $11.941$ & $14.146$ & $7.766$ \\
    $2$  & $\{1,45\}$ & $290\,562$ & $6.9116$ & $9.758$ & $13.460$ & $14.874$ & $8.201$ \\
    $4$  & $\{1,43,44,45\}$ & $581\,124$ & $8.4892$ & $13.700$ & $22.379$ & $17.870$ & $10.139$ \\
    $8$  & $\{1,3,40,42,43,44,45,46\}$ & $1\,162\,248$ & $14.1309$ & $33.861$ & $558.5$ & $29.635$ & $17.005$ \\
    $16$ & $\{1$--$6,8,38$--$46\}$ & $2\,324\,496$ & $42.4301$ & $149.252$ & $1.19\times10^{5}$ & $84.600$ & $50.512$ \\
    \end{tabular}
    \end{ruledtabular}
    \end{table*}

    \begin{table*}[h]
    \caption{Disjoint multi-block downstream accuracy and top-1 agreement.}
    \label{tab:annex_multiblock_downstream}
    \begin{ruledtabular}
    \begin{tabular}{lrrrrrrrr}
    $N$ & LAMBADA & HellaSwag & ARC-e & ARC-c & PIQA & WinoG & BoolQ & top-1 \\
    \colrule
    $1$  & $0.694$ & $0.782$ & $0.847$ & $0.648$ & $0.801$ & $0.714$ & $0.835$ & $0.851$ \\
    $2$  & $0.611$ & $0.775$ & $0.829$ & $0.635$ & $0.801$ & $0.721$ & $0.858$ & $0.819$ \\
    $4$  & $0.651$ & $0.729$ & $0.789$ & $0.596$ & $0.761$ & $0.700$ & $0.846$ & $0.759$ \\
    $8$  & $0.515$ & $0.646$ & $0.693$ & $0.504$ & $0.713$ & $0.665$ & $0.717$ & $0.636$ \\
    $16$ & $0.089$ & $0.423$ & $0.419$ & $0.305$ & $0.608$ & $0.549$ & $0.635$ & $0.430$ \\
    \end{tabular}
    \end{ruledtabular}
    \end{table*}

    \section{Compression accounting}
    \label{annex:compression}

    Writing $P_{\mathrm{base}}=30\,532\,122\,624$ for the base parameter count, $\ell = 623\,120\,640$ for the per-layer count, and $n_{\mathrm{add}}$ for the trainable parameters added,
    \begin{align}
    P_{\mathrm{resulting}} &= P_{\mathrm{base}} - N\ell + n_{\mathrm{add}}, \nonumber\\
    \mathrm{reduction} &= \frac{N\ell - n_{\mathrm{add}}}{P_{\mathrm{base}}}, \qquad
    \mathrm{ratio} = \frac{P_{\mathrm{base}}}{P_{\mathrm{resulting}}}.
    \label{eq:annex_compression}
    \end{align}
    Table~\ref{tab:annex_compression} applies these definitions to every reported configuration; the table is on the total-parameter basis. On the active-parameter basis ($3\,353\,032\,704$ active parameters, $56\,889\,600$ per layer) the reductions are $1.69\%$, $3.39\%$, $6.78\%$, $13.57\%$ and $27.14\%$ for $N=1$, $2$, $4$, $8$ and $16$.
    The operating point we defend is fusion at $N=8$: a $16.33\%$ reduction in stored parameters, a factor of $1.195$.
    The local compression ratios quoted in the main text are defined against the replaced span alone, as $N\ell$ divided by the trainable replacement count.

    \begin{table*}[h]
    \caption{Whole-model compression accounting.}
    \label{tab:annex_compression}
    \begin{ruledtabular}
    \begin{tabular}{lrrrrr}
    Condition & $N$ & Removed & Added & Resulting & Reduction \\
    \colrule
    Single-block, full multi-head QNN & $1$  & $623\,120\,640$    & $145\,281$    & $29\,909\,147\,265$ & $2.0404\%$ \\
    Single-block, MLP      & $1$  & $623\,120\,640$    & $142\,537$    & $29\,909\,144\,521$ & $2.0404\%$ \\
    Single-block, no-circuit   & $1$  & $623\,120\,640$    & $8\,193$      & $29\,909\,010\,177$ & $2.0408\%$ \\
    Identity-skip          & $1$  & $623\,120\,640$    & $0$           & $29\,909\,001\,984$ & $2.0409\%$ \\
    \colrule
    Fusion $N=1$           & $1$  & $623\,120\,640$    & $145\,281$    & $29\,909\,147\,265$ & $2.0404\%$ \\
    Fusion $N=2$           & $2$  & $1\,246\,241\,280$ & $145\,281$    & $29\,286\,026\,625$ & $4.0813\%$ \\
    Fusion $N=4$           & $4$  & $2\,492\,482\,560$ & $145\,281$    & $28\,039\,785\,345$ & $8.1630\%$ \\
    \textbf{Fusion $N=8$}  & $8$  & $4\,984\,965\,120$ & $145\,281$    & $25\,547\,302\,785$ & $\mathbf{16.3265\%}$ \\
    Fusion $N=16$          & $16$ & $9\,969\,930\,240$ & $145\,281$    & $20\,562\,337\,665$ & $32.6534\%$ \\
    \colrule
    Multi-block $N=2$      & $2$  & $1\,246\,241\,280$ & $290\,562$    & $29\,286\,171\,906$ & $4.0808\%$ \\
    Multi-block $N=4$      & $4$  & $2\,492\,482\,560$ & $581\,124$    & $28\,040\,221\,188$ & $8.1616\%$ \\
    Multi-block $N=8$      & $8$  & $4\,984\,965\,120$ & $1\,162\,248$ & $25\,548\,319\,752$ & $16.3231\%$ \\
    Multi-block $N=16$     & $16$ & $9\,969\,930\,240$ & $2\,324\,496$ & $20\,564\,516\,880$ & $32.6463\%$ \\
    \end{tabular}
    \end{ruledtabular}
    \end{table*}

    \section{Statistical resolution and caveats}
    \label{annex:resolution}

    A comparison is called \emph{resolved} when its $95\%$ interval excludes the null; resolved statements below rest on the paired bootstrap where per-document vectors were retained and on the conservative Cauchy--Schwarz bound elsewhere.

    \paragraph{Resolved at 95\%.}
    The single-block full multi-head QNN is better than identity-skipping the same layer (three of three evaluation sets at seed 42) and better than the no-circuit control (nine of nine seed-by-set combinations); every fused span is better than dropping the same span ($15$ of $15$); fusion at $N=1$, $2$ and $4$ is better than the unmodified model (nine of nine, subject to the training-asymmetry confound below); fusion at $N=8$ is worse on WikiText-2 with C4 and \texttt{webmix-recon} unresolved (the parity statement of the main text); fusion at $N=16$ is worse on all three sets; removing the readout at a single depth costs a resolved amount; the single-block full multi-head QNN is worse than the matched MLP at seed 42; disjoint multi-block replacement at $N \geq 4$ is worse than the unmodified model; at seed 42 all three circuit controls are better than the full block on WikiText-2 and C4, the entangler-off control is worse than the matched MLP on WikiText-2 ($+0.022$; C4 unresolved), and the two frozen-circuit controls are worse than the matched MLP on both.

    \paragraph{Directional but not resolved.}
    The full multi-head QNN is worse than the matched MLP at all nine seed-by-set combinations, by $2.9$ to $6.7\%$; only seed 42 is resolved. The frozen-readout ablation costs roughly $2$--$3\%$ perplexity (bound only). Every cross-condition comparison within the fusion ablation is bound-only.

    \paragraph{Bound-only nulls.}
    Register width and upload depth: the single-seed grid cells differ from the deployed configuration by $|\Delta\mathrm{NLL}|<0.017$, none resolved; since these cells also differ in observable-pool composition (caveat 6 below), no width or depth conclusion is drawn.

    Table~\ref{tab:annex_paired} gives the paired-bootstrap comparisons underlying the fusion claims of the main text.

    \begin{table*}[h]
    \caption{%
    Paired bootstrap, fused block against the stated reference, per evaluation set.
    $\Delta\mathrm{NLL}$ is negative when the fused model is better. Intervals are $95\%$ on the perplexity ratio; a comparison is resolved when the interval excludes $1$.}
    \label{tab:annex_paired}
    \begin{ruledtabular}
    \begin{tabular}{llrrlc}
    Comparison & Eval set & $\Delta\mathrm{NLL}$ & Ratio & $95\%$ CI & Resolved \\
    \colrule
    Fusion $N=1$ vs.\ baseline  & WikiText-2 & $-0.0628$ & $0.939$ & $[0.936,\,0.943]$ & yes \\
    Fusion $N=1$ vs.\ baseline  & C4         & $-0.0485$ & $0.953$ & $[0.951,\,0.955]$ & yes \\
    Fusion $N=1$ vs.\ baseline  & \texttt{webmix-recon} & $-0.0668$ & $0.935$ & $[0.929,\,0.942]$ & yes \\
    Fusion $N=2$ vs.\ baseline  & WikiText-2 & $-0.0499$ & $0.951$ & $[0.948,\,0.955]$ & yes \\
    Fusion $N=2$ vs.\ baseline  & C4         & $-0.0424$ & $0.958$ & $[0.956,\,0.961]$ & yes \\
    Fusion $N=2$ vs.\ baseline  & \texttt{webmix-recon} & $-0.0565$ & $0.945$ & $[0.939,\,0.952]$ & yes \\
    Fusion $N=4$ vs.\ baseline  & WikiText-2 & $-0.0422$ & $0.959$ & $[0.955,\,0.962]$ & yes \\
    Fusion $N=4$ vs.\ baseline  & C4         & $-0.0278$ & $0.973$ & $[0.970,\,0.975]$ & yes \\
    Fusion $N=4$ vs.\ baseline  & \texttt{webmix-recon} & $-0.0356$ & $0.965$ & $[0.958,\,0.972]$ & yes \\
    Fusion $N=8$ vs.\ baseline  & WikiText-2 & $+0.0087$ & $1.009$ & $[1.005,\,1.013]$ & yes \\
    Fusion $N=16$ vs.\ baseline & WikiText-2 & $+0.1342$ & $1.144$ & $[1.137,\,1.150]$ & yes \\
    Fusion $N=16$ vs.\ identity-skip & WikiText-2 & $-0.2794$ & $0.756$ & $[0.752,\,0.761]$ & yes \\
    \end{tabular}
    \end{ruledtabular}
    \end{table*}

    \subsection{Caveats and confounds}
    \label{annex:caveats}
    \begin{enumerate}
    \setlength{\itemsep}{0pt}\setlength{\parskip}{0pt}
    \item \textbf{Training asymmetry.} The replacement is trained for $20\,000$ steps on \texttt{webmix\_calib} and the base model is not; gains over the unmodified model are not a capability claim, since identity-skip is untrained and no fine-tuned base model was run.
    \item \textbf{\texttt{webmix-recon} is not held out} (in-corpus, descriptive only).
    \item \textbf{Validation perplexity is a selection metric} (minimum of $40$ measurements) and carries the corresponding optimism; no cross-condition claim rests on it.
    \item \textbf{The BoolQ drop is consistent with fine-tuning, not architecture:} the full block and the matched MLP reach $0.83$ against $0.885$ for the base.
    \item \textbf{Register-width cells} differ in observable-pool composition as well as qubit count; no width comparison is made.
    \item \textbf{Per-document likelihoods} were retained for twelve conditions only; every other comparison rests on the conservative bound.
    \item \textbf{Circuit controls are largely single-seed} (Appendix~\ref{annex:controls}); only the entangler-off control has a second seed.
    \item \textbf{Beyond the headline configurations the campaign is single-seed.}
    \end{enumerate}

    \textbf{Anomalies, disclosed rather than smoothed over.} One seed-7 cell shows an anomalous word-level perplexity ($38.128$ against a token perplexity of $9.387$); word-level perplexity is excluded from every aggregate and comparison. Some evaluation curves show large transient spikes (the $n_q=4$ three-pass cell reaches $7.768$ at step $20\,000$ against a best of $7.0175$). One block of the $N=8$ multi-block configuration converges to a negative residual gate, $\alpha=-0.0339$; undiagnosed. The $N=12$ fused span was trained but never benchmarked, and is not reported.

    \textbf{What this evidence does not establish.} No quantum advantage; no claim that a quantum module is \emph{necessary}, only that, at a sensitive block, the one built performs work beyond that of the surrounding classical components; no preservation of all model capabilities (the position-wise structure removes cross-token mixing at every replaced depth); no qubit-count or depth-scaling law; and no comparison against practical compression pipelines; and, given the circuit controls of Appendix~\ref{annex:controls}, no claim that the single-block recovery is specific to trained or entangled circuits.

    \end{document}